\documentclass[a4paper,11pt]{article}
\usepackage[numbers,square,sort&compress,sectionbib]{natbib}
\usepackage{chapterbib}
\usepackage{graphicx}
\usepackage{hyperref}

\usepackage{setspace}
\usepackage{pslatex}
\usepackage{amsmath}
\usepackage{bm}
\usepackage{braket}
\numberwithin{equation}{section}
\numberwithin{table}{section}
\numberwithin{figure}{section}
\usepackage[font=small,labelfont=bf]{caption}
\usepackage{multirow}
\usepackage[toc,page]{appendix}
\usepackage[font=small,labelfont=bf]{caption}
\usepackage{placeins}
\usepackage[top=2.5cm, bottom=2.5cm, left=3.8cm, right=2.5cm]{geometry}
\usepackage{changepage}
\usepackage{titlepic}
\usepackage{tikz}
\usepackage{amssymb}
\usepackage{pdfpages}
\usetikzlibrary{calc,shapes,arrows}
\doublespace

\title{\begin{huge}
\textbf{Theory of electron and phonon transport in nano and molecular quantum devices}\\ \vspace*{1cm}
\end{huge} \Large {Design strategies for molecular electronics and thermoelectricity\\}}
\author{\\Dr. Hatef Sadeghi \\ Quantum Technology Center, Lancaster University, Lancaster, UK\\\textit{h.sadeghi@lancaster.ac.uk; hatef.sadeghi@gmail.com}}
\date{\today}
\begin{document}
\maketitle
  \tableofcontents
  \section{Introduction: Molecular electronics}

\begin{figure}[htbp] 
   \centerline{\includegraphics[width=1\textwidth]{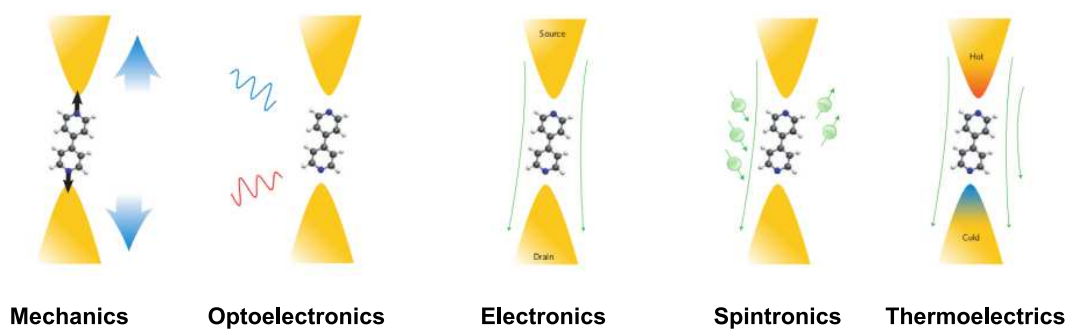}}
  \caption{\textbf{Molecular electronic active area of research}\cite{Aradhya2013}}
  \label{intfig}
\end{figure}

The idea of using single molecules as building blocks to design and fabricate molecular electronic components has been around for more than 40 years \cite{natnano2013v}, but only recently it has attracted huge scientific interest to explore their unique properties and opportunities. Molecular electronics including self-assembled monolayers \cite{revselfasebleded} and single-molecule junctions \cite{Aradhya2013} are of interest not only for their potential to deliver logic gates \cite{Sangtarash2015}, sensors\cite{Sadeghi2014a}, and memories \cite{Prodromakis2012} with ultra-low power requirements and sub-10-nm device footprints, but also for their ability to probe room-temperature quantum properties at a molecular scale such as quantum interference \cite{C4CS00203B} and thermoelectricity \cite{sadeghi2015oligoyne}. There are five main area of research in molecular-scale electronics \cite{Aradhya2013} namely: Molecular mechanics, molecular optoelectronics, molecular electronics, molecular spintronics and molecular thermoelectrics as shown in figure \ref{intfig} in which studying the electronic and phononic transport properties of the junction is the central basis toward junction characterization for a wide range of the applications. 

By studying electron and phonon transport across a junction consisting of two or more electrodes connected to a single or a few hundred molecules, one could study all phenomenon shown in figure \ref{intfig} from mechanical properties of the junction to electronic and thermoelectrics. For example, when a single molecule is attached to metallic electrodes, de Broglie waves of electrons entering the molecule from one electrode and leaving through the other form complex interference patterns inside the molecule. These patterns could be utilize to optimize the single-molecule device performance \cite{Sangtarash2015,Geng2015}. Furthermore, recently their potential for removing heat from nanoelectronic devices (thermal management) and thermoelectrically converting waste heat into electricity has also been recognised \cite{sadeghi2015oligoyne}. Indeed, electrons passing through single molecules have been demonstrated to remain phase coherent, even at room temperature. In practice, the task of identifying and harnessing quantum effects is hampered because transport properties are strongly affected by the method used to anchor single molecules to electrodes. My aim in this paper is to review the theoretical and mathematical techniques to treat electron and phonon transport in nano and molecular scale junctions leading to models of their physical properties. This helps not only to understand the experimental observations but also provides a vital design tool to develop strategies for molecular electronic building blocks, thermoelectric device and sensors. 

\section{Transport in molecular scale} \label{Transport in molecular scale}

\begin{figure}[htbp] 
   \centerline{\includegraphics[width=0.6\textwidth]{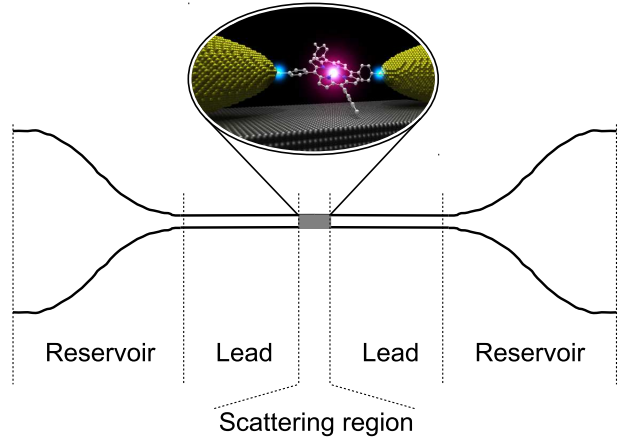}}
  \caption{\textbf{A scattering region is connected to the reservoirs trough ballistic leads.} Reservoirs have slightly different electrochemical potentials to drive electrons from the left to the right lead. All inelastic relaxation process take place in the reservoirs and transport in the leads are ballistic.}
  \label{box0}
\end{figure}

My focus in this paper is on reviewing the methods used to model electron and phonon transport in nano and molecular scale systems. Any device consists of two or more electrodes (leads) connected to a scattering region (figure \ref{box0}). The electrodes are perfect waveguides where electrons and phonons transmit without any scattering. The main scattering occurs either at the junction to the leads or inside the scattering region. The goal is to understand electrical and vibrational properties of nano and molecular junctions where nanoscale scatter or molecules are the bridge between the electrodes with or without surroundings, such as an electric field (gate and bias voltages or local charge), a magnetic field, a laser beam or a molecular environment (water, gases, biological spices, donors and acceptors, etc). In principle, the molecule could be coupled to the electrodes with a weak or strong coupling strength. However, in most cases the coupling is weak. There are different approaches to study the electronic and vibrational properties of the junctions \cite{Datta2005b} though, my focus in this paper is mostly on the Green's function formalism and partially the master equation approach.

Here, I will begin with the Schr\"{o}dinger equation and try to relate it to the physical description of matter at the nano and molecular scale. Then I will discuss the definition of the current using the time-dependent Schr\"{o}dinger equation and introduce tight binding description of the quantum system. The scattering theory and non-equilibrium Green's function method are discussed and different transport regimes (on and off resonances) are considered. One dimensional system and a more general multi-channel method are derived to calculate transmission coefficient $T(E)$ in a molecular junction for electrons (phonons) with energy $E$ ($\hbar \omega$) traversing from one electrode to another. I then briefly discuss the master equation method to model transport in the Coulomb and Franck-Condon blockade regimes. I follow with a discussion about physical interpretation of a quantum system and different techniques used to model the experiment.

\subsection{Schr\"{o}dinger equation} \label{Schrodinger equation} 

The most general Schr\"{o}dinger equation \cite{PhysRevSchr} describes the evolution of the physical properties of a system in time and was proposed by the Austrian physicist Erwin Schr\"{o}dinger in 1926 as:

\begin{equation} \label{sch-eqT}
i\hbar {\partial \over \partial t}\Psi(r,t) = \hat{H} \Psi(r,t)
\end{equation}
where $i$ is $\sqrt{-1}$, $\hbar$ is the reduced Planck constant ($h / {2\pi}$), $\Psi$ is the wave function of the quantum system, and $\hat{H}$ is the Hamiltonian operator which characterizes the total energy of any given wave function. For a single particle moving in an electric field, the non-relativistic Schr\"{o}dinger equation reads as:

\begin{equation} \label{sch-eqX}
i\hbar {\partial \over \partial t}\Psi(r,t) = [{-\hbar^2 \over {2m}} \bigtriangledown ^2 + V(r,t)] \Psi(r,t)
\end{equation}
If we write the wavefunction as a product of spatial and temporal terms: $\Psi(r,t) = \psi(r)\theta(t)$, the Schr\"{o}dinger equation become two ordinary differential equations:

\begin{equation} \label{sch-eqt0}
\frac{1}{\theta(t)}\frac{d}{dt}\theta(t) = -\frac{iE}{\hbar}
\end{equation}
and
\begin{equation} \label{sch-eqx0}
\hat{H}\psi(r) = E\psi(r)
\end{equation}
where ${\hat{H} = \frac{-\hbar^2}{2m}} \bigtriangledown^2+V(r)$. The solution of equation \ref{sch-eqt0} could be written as: $\theta(t) = e^{-iEt/\hbar}$. The amplitude of $\theta(t)$ does not change with time and therefore the solutions $\theta(t)$ are purely oscillatory. The total wave function 

\begin{equation} \label{sol-sch-eq}
\Psi(r,t) = \psi(r)e^{-iEt/\hbar}
\end{equation}
differs from $\psi(r)$ only by a phase factor of constant magnitude and the expectation value $\vert \Psi(r,t) \vert^2$ is time-independent. Of course \ref{sol-sch-eq} is a particular solution of time-dependent Schr\"{o}dinger equation. The most general solution is a linear combination of these particular solutions as:

\begin{equation} \label{gen-sol-sch-eq}
\Psi(r,t) = \sum_i \phi_i e^{-iE_it/\hbar} \psi_i(r)
\end{equation}
In time independent problems only the spatial part needs to be solved since the time dependent phase factor in \ref{sol-sch-eq} is always the same. Equation \ref{sch-eqx0} is called time-independent Schr\"{o}dinger equation and it is an eigenvalue problem where $E$'s are eigenvalues of the Hamiltonian $\hat{H}$. Since the Hamiltonian is a Hermitian operator, the eigenvalues $E$ are real. $\psi(r)$ describes the standing wave solutions of the time-dependent equation, which are the states with definite energy called "stationary states" or "energy eigenstates" in physics and "atomic orbitals" or "molecular orbitals" in chemistry. 

The Schr\"{o}dinger equation must be solved subject to appropriate boundary conditions. Since the electrons are fermions, the solution must satisfy the Pauli exclusion principle and wavefunction $\psi$ must be well behaved everywhere. The Schr\"{o}dinger equation can be solved analytically for a few small systems such as  the hydrogen atom. However, this is too complex to be solved in most cases even with the best supercomputers available today, so some approximations are needed \cite{Engel2011} such as the Born-Oppenhaimer approximation to decouple the movement of the electrons and the nuclei; density functional theory (DFT) to describe the electron - electron interactions and pseudopotentials to treat the nuclei and the core electrons except those in the valence band. These methods are well-known and are described in \cite{Engel2011} and breifley discussed in the next section. To describe the transport through the molecules or nanoscale matters, one needs to build a simple tight-binding Hamiltonian using Huckel parameters or use DFT to construct mean-field Hamiltonian.

To reduce the size of the Hamiltonian, it is appropriate to define the idea of the basis functions where

\begin{equation} \label{Xsol-sch-eq}
\Psi(r) = \sum_i \phi_i \psi_i(r)
\end{equation}
The wavefunction then can be represented by a column vector $\vert \phi \rangle$ consisting of the expansion coefficients $\phi_i$. The time-independent Schr\"{o}dinger equation could be written as a matrix equation:

\begin{equation} \label{mat-sch-eq}
[H]\vert \phi \rangle = E[S]\vert \phi \rangle
\end{equation}
where 
\begin{equation} \label{S-eq}
S_{ij} = \langle i \vert j \rangle = \int dr \psi_j^*(r)\psi_i(r)
\end{equation}
and
\begin{equation} \label{H-eq}
H_{ij} =  \langle i \vert H \vert j \rangle = \int dr \psi_j^*(r)H\psi_i(r)
\end{equation}
The evaluation of these integrals is the most time-consuming step, but once $[H]$ and $[S]$ are obtained, the eigenvalues $E_n$ and eigenvectors $\phi_n$ are easily calculated. 
If $\langle i \vert$ and $\vert j \rangle$ are orthogonal then $S_{ij} = \delta_{ij}$ where $\delta_{ij}$ is the Kronecker delta ($\delta_{ij} = 1$ if $i=j$ and $\delta_{ij} = 0$ if $i \neq j$). 

\subsection{Density functional theory (DFT)}

\begin{figure}[htbp] 
   \centerline{\includegraphics[width=1\textwidth]{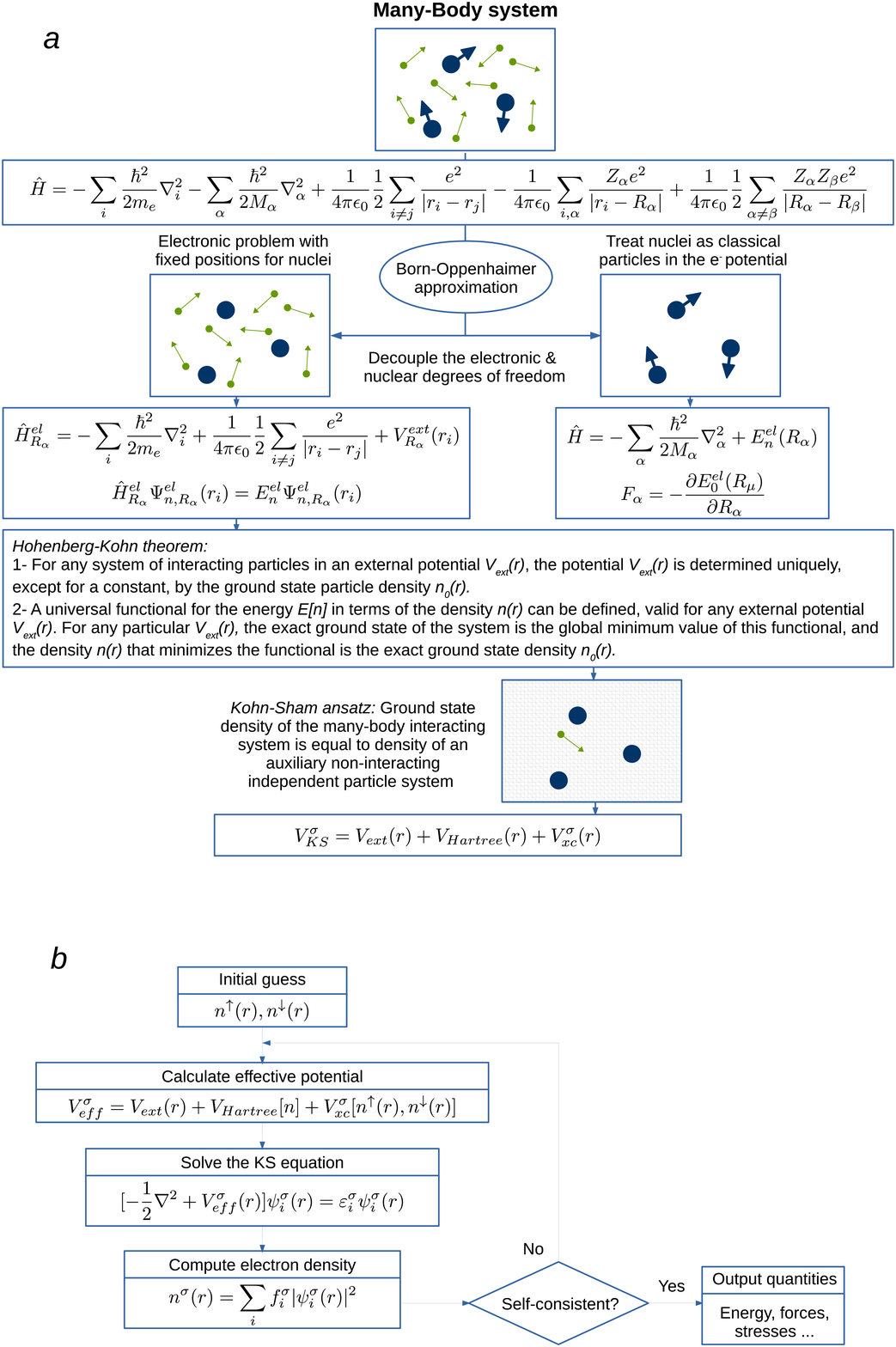}}
  \caption{\textbf{From many-body problem to density functional theory DFT.} (a) Born-Oppenheimer approximation, Hohenberg-Kohn theorem and Kohn-Sham ansatz, (b) Schematic of the DFT self-consistency process.}
  \label{box14b}
\end{figure}

In order to understand the behaviour of molecular electronic devices, it is necessary to possess a reliable source of structural and electronic information. A solution to the many body problem has been sought by many generations of physicists. The task is to find the eigenvalues and eigenstates of the full Hamiltonian operator of a system consisting of nuclei and electrons as shown in figure \ref{box14b}. Since this is not practically possible for the systems bigger than a few particles, some approximations are needed. The atomic masses are roughly three orders of magnitudes bigger than the electron mass, hence the Born-Oppenheimer approximation \cite{Engel2011} can be employed to decouple the electronic wave function and the motion of the nuclei. In other words we solve the Schr\"{o}dinger equation for the electronic degrees of freedom only. Once we know the electronic structure of a system we can calculate classical forces on the nuclei and minimize these forces to find the ground-state geometry (figure \ref{box14b}a). 

Once the Schr\"{o}dinger equation was solved, the wavefunction is known and all physical quantities of intereste could be calculated. Although the Born-Oppenheimer approximation decouple the electronic wave function and the motion of the nuclei, the electronic part of the problem has reduced to many interacting particles problem which even for modest system sizes i.e. a couple of atoms, its diagonalization is practically impossible even on a modern supercomputer. The virtue of density functional theory DFT \cite{Engel2011,harrison2003introduction} is that it expresses the physical quantities in terms of the ground-state density and by obtaining the ground-state density, one can in principle calculate the ground-state energy. However, the exact form of the functional is not known. The kinetic term and internal energies of the interacting particles cannot generally be expressed as functionals of the density. The solution is introduced by Kohn and Sham in 1965. According to Kohn and Sham, the original Hamiltonian of the many body interacting system can be replaced by an effective Hamiltonian of non-interacting particles in an effective external potential, which has the same ground-state density as the original system as illustrated in figure \ref{box14b}a. The difference between the energy of the non-interacting and interacting system is referred to the exchange 
correlation functional (figure \ref{box14b}a).

\textbf{Exchange and correlation energy:} There are numerous proposed forms for the exchange and correlation energy $V_{xc}$ in the literature \cite{Engel2011,harrison2003introduction}. The first successful - and yet simple - form was the Local Density Approximation (LDA) \cite{harrison2003introduction}, which depends only on the density and is therefore a local functional. Then the next step was the Generalized Gradient Approximation (GGA) \cite{harrison2003introduction}, including the derivative of the density. It also contains information about the neighborhood and therefore is semi-local. LDA and GGA are the two most commonly used approximations to the exchange and correlation energies in density functional theory. There are also several other functionals, which go beyond LDA and GGA. Some of these functionals are tailored to fit specific needs of basis sets used in solving the Kohn-Sham equations and a large category are the so called hybrid functionals (eg. B3LYP, HSE and Meta hybrid GGA), which include exact exchange terms from Hartree-Fock. One of the latest and most universal functionals, the Van der Waals density functional (vdW-DF), contains non-local terms and has proven to be very accurate 
in systems where dispersion forces are important. 

\textbf{Pseudopotentials:} Despite all simplifications shown in \ref{box14b}, in typical systems of molecules which contain many atoms, the calculation is still very large and has the potential to be computationally expensive. In order to reduce the number of electrons, one can introduce pseudopotentials which effectively remove the core electrons from an atom. The electrons in an atom can be split into two types: core and valence, where core electrons lie within filled atomic shells and the valence electrons lie in partially filled shells. Together with the fact that core electrons are spatially localized about the nucleus, only valence electron states overlap when atoms are brought together so that in most systems only valence electrons contribute to the formation of molecular orbitals. This allows the core electrons to be removed and replaced by a pseudopotential such that the valence electrons still feel the same screened nucleon charge as if the core electrons were still present. This reduces the number of electrons in a system dramatically and in turn reduces the time and memory required to calculate properties of molecules that contain a large number of electrons. Another benefit of pseudopotentials is that they are smooth, leading to greater numerical stability. 

\textbf{Basis Sets:} For a periodic system, the plane-wave basis set is natural since it is, by itself, periodic. However, since we need to construct a tight-binding Hamiltonian, we need to use localised basis sets discussed in the next section, which are not implicitly periodic. An example is a Linear Combination of Atomic Orbital (LCAO) basis set which are constrained to be zero after some defined cut-off radius, and are constructed from the orbitals of the atoms.

To obtain a ground state mean-field Hamiltonian from DFT, the calculation is started by constructing the initial atomic configuration of the system. Depending on the applied DFT implementation, the appropriate pseudopotentials for each element which can be different for every exchange-correlation functional might be needed. Furthermore, a suitable choice of the basis set has to be made for each element present in the calculation. The larger the basis set, the more accurate our calculation - and, of course, the longer it will take. With a couple of test calculations we can optimize the accuracy and computational cost. Other input parameters are also needed that set the accuracy of the calculation such as the fineness and density of the $k$-grid points used to evaluate the integral(\cite{harrison2003introduction,Soler2002}). Then an initial charge density assuming no interaction between atoms is calculated. Since the pseudopotentials are known this step is simple and the total charge density will be the sum of the atomic densities. 

The self-consistent calculation \cite{harrison2003introduction}(figure \ref{box14b}b) starts by calculating the Hartree potential and the exchange correlation potential. Since the density is represented in real space, the  Hartree potential is obtained by solving the Poisson equation with the multi-grid or fast Fourier-transform method, and the exchange-correlation potential is obtained. Then the Kohn-Sham equations are solved and a new density is obtained. This self-consistent iterations end when the necessary convergence criteria are reached such as density matrix tolerance. Once the initial electronic structure of a system obtained, the forces on the nucleis could be calculated and a new atomic configuration to minimize these forces obtained. New atomic configuration is new initial coordinate for self-consistent calculation. This structural optimization is controlled by the conjugate gradient method for finding the minimal ground state energy and the corresponding atomic configuration \cite{harrison2003introduction}. From the obtained ground state geometry of the system, the ground state electronic properties of the system such as total energy, binding energies between different part of the system, density of states, local density of states, forces, etc could be calculated. It is apparent that the DFT could potentially provide an accurate description of the ground state properties of a system such as total energy, binding energy and geometrical structures. However, DFT has not been originally designed to describe the excited state properties and therefore all electronic properties related to excited states are less accurate within DFT. If the LCAO basis is used, the Hamiltonian and overlap matrices used within the scattering calculation could be extracted.

\subsection{Tight-Binding Model} \label{Tight-binding Hamiltonian}

By expanding the wavefunction over a finite set of the atomic orbitals, the Hamiltonian of the system can be written in a tight-binding model. The main idea is to represent the wave function of a particle as a linear combination of some known localized states. A typical choice is to consider a linear combination of atomic orbitals (LCAO). For a periodic system where the wave-function is described by a Bloch function, equation \ref{mat-sch-eq} could be written as

\begin{equation} \label{lcao-eq0}
\sum_{\beta,c'} H_{\alpha,c;\beta,c'} \phi_{\beta,c'} = E \sum_{\beta,c'} S_{\alpha,c;\beta,c'} \phi_{\beta,c'}
\end{equation}
where $c$ and $c'$ are the neighbouring identical cells containing states $\alpha$ and

\begin{equation} \label{lcao-eq1}
H_{\alpha,c;\beta,c'} = H_{\alpha,\beta}(R_c-R_{c'})
\end{equation}
and

\begin{equation} \label{lcao-eq2}
\phi_{\beta,c} = \phi_\beta e^{ik.R_c}
\end{equation}
The equation \ref{lcao-eq0} could be written as

\begin{equation} \label{lcao-eq3}
\sum_{\beta} H_{\alpha\beta}(k) \phi_{\beta} = E \sum_{\beta} S_{\alpha\beta}(k) \phi_{\beta}
\end{equation}
where

\begin{equation} \label{lcao-eq4}
H_{\alpha\beta}(k) = \sum_{c'} H_{\alpha\beta}(R_c-R_{c'}) e^{ik(R_c-R_{c'})} 
\end{equation}
and

\begin{equation} \label{lcao-eq5}
S_{\alpha\beta}(k) = \sum_{c'} S_{\alpha\beta}(R_c-R_{c'}) e^{ik(R_c-R_{c'})} 
\end{equation}
More generally, the single-particle tight-binding Hamiltonian in the Hilbert space formed by $\vert R_\alpha \rangle$ could be written as:

\begin{equation} \label{TB-H}
H = \sum_\alpha (\varepsilon_\alpha + e V_\alpha) \vert \alpha \rangle \langle \alpha \vert + \sum_{\alpha \beta} \gamma_{\alpha \beta} \vert \alpha \rangle \langle \beta \vert
\end{equation}
where $\varepsilon_\alpha$ is the corresponding on-site energy of the state $\vert \alpha \rangle$, $V_\alpha$ is the electrical potential and the $\gamma_{\alpha \beta}$ is the hopping matrix element between states $\vert \alpha \rangle$ and $\vert \beta \rangle$. For conjugated hydrocarbon systems, the energies of molecular orbitals associated with the pi electrons could be determined by a very simple LCAO molecular orbitals method called Huckel molecular orbital method (HMO). Therefore, a simple TB description of the system could be conduct just by assigning a Huckel parameter for on-site energy $\varepsilon_\alpha$ of each atom in the molecule connected to the nearest neighbours with a single Huckel parameter for hopping matrix element $\gamma_{\alpha \beta}$. Obviously, more complex TB models could be made using HMO by taking second, third, forth or more nearest neighbours hopping matrix element into account.

\subsubsection{One dimensional (1D) infinite chain} \label{One dimensional (1D) infinite chain}

As an example, a single-orbital orthogonal nearest neighbour tight binding Hamiltonian of an infinite linear chain of hydrogen atoms shown in figure \ref{box2} with on-site energy $\langle j \vert H \vert j \rangle = \varepsilon_0$ and the hopping matrix element $\langle j \vert H \vert j\pm 1 \rangle = \langle j\pm 1 \vert H \vert j \rangle  = -\gamma$ could be written as:
\begin{figure}[htbp] 
   \centerline{\includegraphics[width=0.8\textwidth]{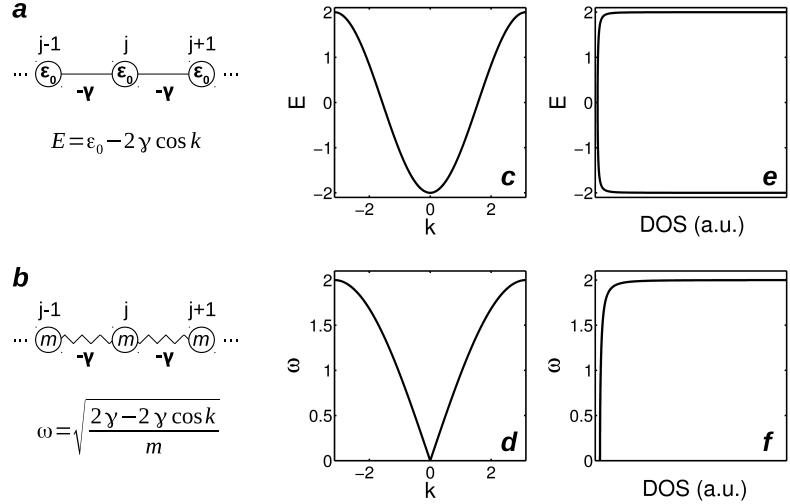}}
  \caption{\textbf{One dimensional (1D) infinite chain}. (a) hydrogen atoms in an infinite chain with one orbital per atom, (b) 1D balls and springs, (c,d) electronic and phononic band structures and (e,f) density of states (DOS) for a and b.}
  \label{box2}
\end{figure}
\begin{equation} \label{HTB1D-H}
H = \sum_j \varepsilon_0 \vert j \rangle \langle j \vert - \sum_{j,j+1} \gamma \vert j \rangle \langle j+1 \vert - \sum_{j-1,j} \gamma \vert j-1 \rangle \langle j \vert
\end{equation}
Therefore the Schr\"{o}dinger equation reads

\begin{equation} \label{TB1D-H}
\varepsilon_0 \phi_j - \gamma \phi_{j-1} - \gamma \phi_{j+1} = E \phi_j
\end{equation}
where $-\infty < j < +\infty$. The solution of this equation could be obtained using the Bloch function as

\begin{equation} \label{TB1D-w}
\vert \psi_k \rangle = \frac{1}{\sqrt{N}} \sum_{j} e^{ikja_0} \vert j \rangle
\end{equation}
and 

\begin{equation} \label{TB1D-E}
E(k) = \varepsilon_0 -2\gamma cos(ka_0)
\end{equation}
where $-\pi/a_0<k<\pi/a_0$ in the first Brillouin zone. Equation \ref{TB1D-E} is called a dispersion relation ($E-k$) or electronic bandstructure of a 1D chain. Since $-1<cos (ka_0)<1$, hence $\varepsilon_0-2\gamma<E<\varepsilon_0+2\gamma$; therefore the bandwidth is $4\gamma$. The density of states (DOS) could be calculated from:

\begin{equation} \label{dos}
D(E) = \sum_i \delta(E-\varepsilon_i)
\end{equation}
where $\varepsilon_i$ is the eigenvalues of a system and $\delta$ is Kronecker delta. Figure \ref{box2}a shows the band structure and density of states for a 1D chain.

I have yet discussed the electronic properties of a quantum system e.g. 1D chain. Now consider a chain of the atoms with mass $m$ connected to each other with the springs with spring-constant $K=-\gamma$ as shown in figure \ref{box2}. In one hand, the derivative of the energy with respect to the position of the atoms describe the forces in the system ($F=-\frac{\partial}{\partial x}U$). On the other hand, from Newton's second law $F=-m\frac{d^2 x}{d t^2}$. Using the harmonic approximation method the Schr\"{o}dinger-like equation could be written as:

\begin{equation} \label{Dph-sch}
-m\frac{d^2 x_n}{d t^2} = -K[2x_n-x_{n-1}-x_{n+1}]
\end{equation}
Similar to what was discussed above, using $x_n(t) = A e^{i(kn-\omega t)}$, equation \ref{Dph-sch} reads $-m\omega^2 = -K[2-e^{-ik}-e^{ik}]$ and therefore the phononic dispersion relation is obtained as

\begin{equation} \label{1Dph-w}
\omega(k) = \sqrt{\frac{2\gamma-2\gamma cos k}{m}}
\end{equation}
Comparing the equation \ref{TB1D-E} and \ref{1Dph-w}, it is apparent that the equation \ref{1Dph-w} could be written by changing the $E \rightarrow m\omega^2$ and $\varepsilon_0 \rightarrow 2\gamma$ in the equation \ref{TB1D-E}.  $\varepsilon_0 =2\gamma$ is the negative of the sum of all off-diagonal terms of the 1D chain TB Hamiltonian in which make sense to satisfy translational invariance. The general Schr\"{o}dinger equation for phonons could be written as 

\begin{equation} \label{1Dph-w1}
\omega^2 \psi= D \psi
\end{equation}
This is very similar to the equation \ref{mat-sch-eq}, where $E \rightarrow \omega^2$, and the dynamical matrix $D=-K/M$ where $M$ is the mass matrix, $K_{ij}$ could be calculated from the force matrix and $K_{ii}=\sum_{i \neq j} K_{ij}$.

\subsubsection{One dimensional (1D) finite chain and ring} \label{One dimensional (1D) finite chain and ring}
\begin{figure}[htbp] 
   \centerline{\includegraphics[width=0.7\textwidth]{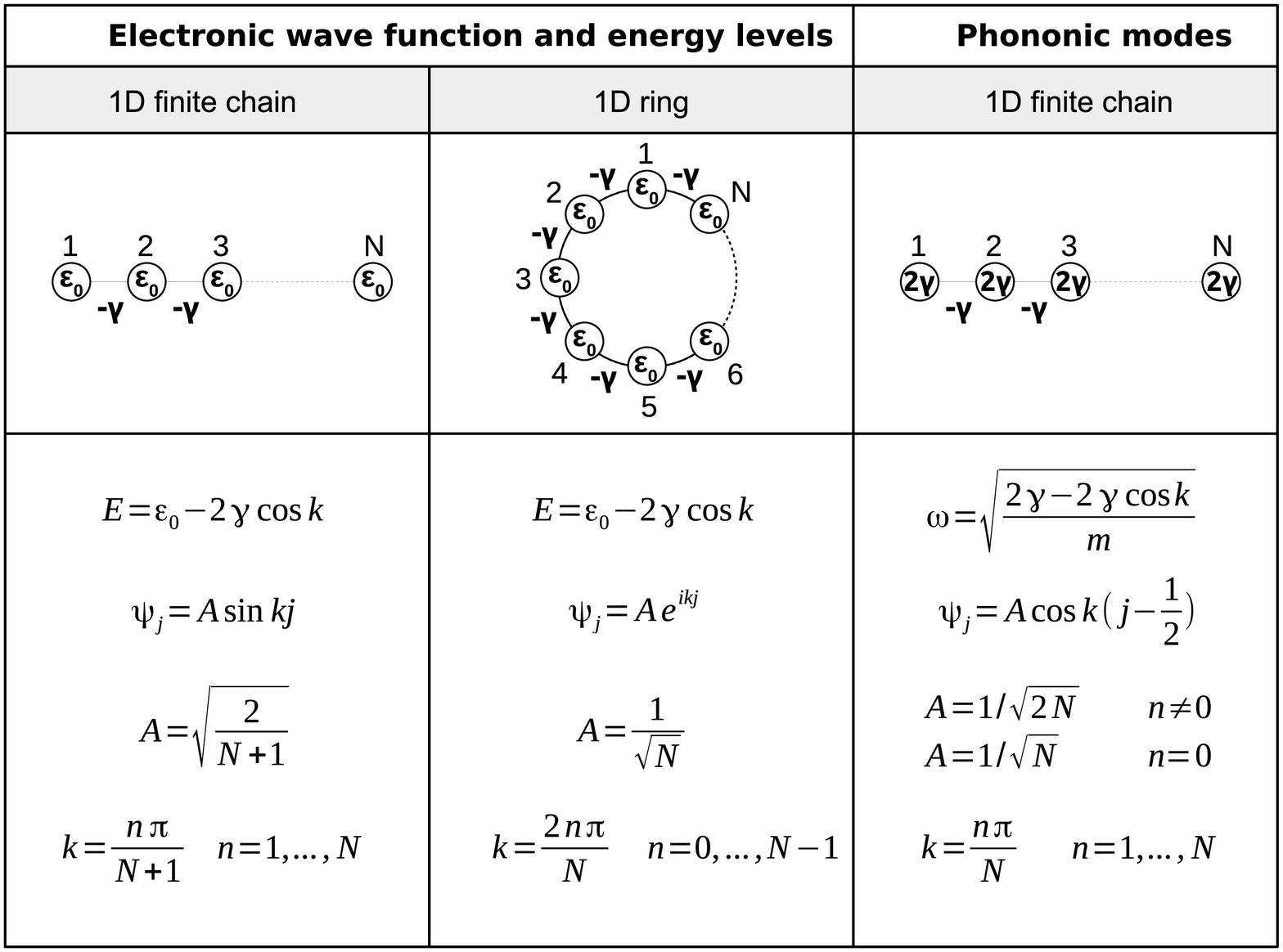}}
  \caption{\textbf{1D finite chain and ring}. The energy levels and corresponding wave functions or orbitals for a 1D finite chain and ring. The phononic mode for a finite chain of balls and springs with mass $m$.}
  \label{box1}
\end{figure}
To analyse the effect of the different boundary conditions in the solution of the Schr\"{o}dinger equation, I consider three examples shown in figure \ref{box1}. Consider a 1D finite chain of $N$ atoms. As a consequence of introducing the boundary condition at the two ends of the chain, the energy levels and states are no longer (continuous) in the range of $\varepsilon_0-2\gamma<E<\varepsilon_0+2\gamma$; instead there are discrete energy levels and corresponding states in this range. The differences in the allowed energy levels between a 1D finite chain and a 1D ring demonstrates that small changes in the system significantly affect the energy levels and corresponding orbitals. This is more important where few number of atoms investigated e.g. the molecules, so two very similar molecule could show different electronic properties. 

\subsubsection{Two dimensional (2D) square and hexagonal lattices} \label{Two dimensional square and hexagonal lattices}
\begin{figure}[htbp] 
   \centerline{\includegraphics[width=1.1\textwidth]{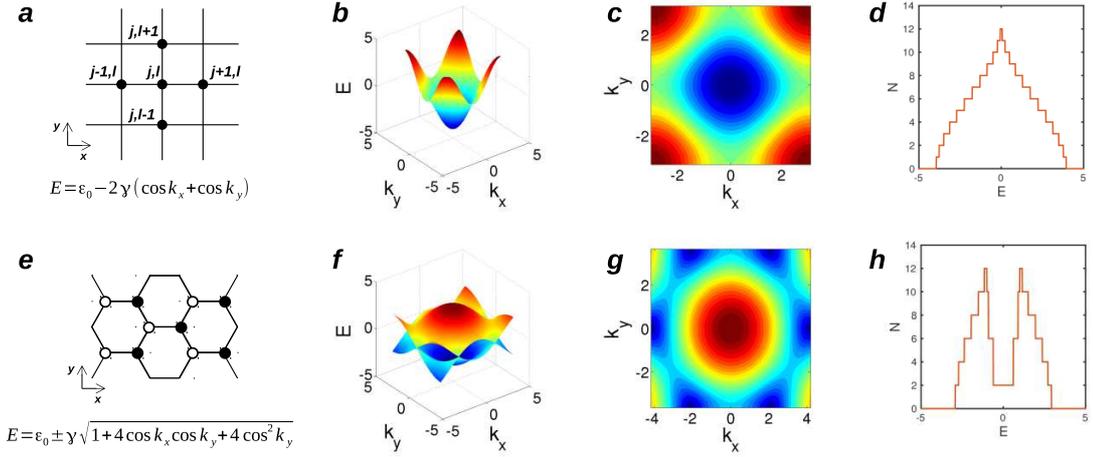}}
  \caption{\textbf{Two dimensional square and hexagonal lattices}. Lattice geometry of (a) square and (e) hexagonal lattices, the bandstructure of (b,c) square and (f,g) hexagonal lattices and the number of conduction channels in (d) square and (h) hexagonal lattices.}
  \label{box3}
\end{figure}

Using the TB Hamiltonian of a 1D chain, I calculated its band-structure and density of states. Now let's consider two most used 2D lattices: a square lattice where the unit-cell consist of one atom is connected to the first nearest neighbour in two dimensions (figure \ref{box3}a) and a hexagonal lattice where a unit cell consist of two atoms is connected to the neighbouring cells in which first (second) atom in a cell is only connected to the second (first) atom in any first nearest neighbour cell (figure \ref{box3}b). The TB Hamiltonian and corresponding band-structure could be calculated \cite{Datta2005b} using the equation \ref{TB-H} and the Bloch wave function has the form of $Ae^{ik_xj+ik_yl}$ as shown in figure \ref{box3}.

Figures \ref{box3}b,c,f,g show the bandstructure of square and hexagonal lattices. Furthermore, the number of conduction channels could be calculated as shown in figures \ref{box3}d,h using the method described in section \ref{Generalized model to calculate T(E)}. The number of channels has a maximum in the middle of the band for a square lattice, whereas for a hexagonal lattice, there are fewer open channels (e.g. only two for graphene) in the middle of the band.

\subsection{Current carried by a Bloch function} \label{2D square and hexagonal lattices}
The time evolution of the density matrix $\rho_t = \vert \psi_t\rangle \langle \psi_t \vert$ allows us to obtain current associated with a particular quantum state $\vert \psi_t \rangle$. Using the time-dependent Schr\"{o}dinger equation \ref{sch-eqT}, I define

\begin{equation} \label{current-psi}
I = \frac{d}{dt}\vert \psi_t \rangle \langle \psi_t \vert = \frac{1}{i\hbar} [H\vert \psi_t\rangle \langle \psi_t \vert-\vert \psi_t\rangle \langle \psi_t \vert H]
\end{equation}
By expanding $\vert \psi_t \rangle$ over orthogonal basis $\vert j \rangle$ equation \ref{current-psi} could be written as:

\begin{equation} \label{current-psi2}
\frac{d\rho_t}{dt}= \frac{1}{i\hbar} [\sum_{jj'} H\vert j \rangle \langle j' \vert \psi_j\psi_{j'}^* -\sum_{jj'}\vert j \rangle \langle j' \vert H \psi_j\psi_{j'}^*]
\end{equation}
For a 1D infinite chain with the Hamiltonian of the form of \ref{HTB1D-H}, the rate of change of charge $I_l = {d\rho_t^l}/{dt}$ at site $l$ could be obtained by calculating the expectation value of both side of \ref{current-psi2} over the state $\vert l \rangle$

\begin{equation} \label{current-psi3}
\frac{d\rho_t^l}{dt}= \frac{1}{i\hbar} [\sum_{jj'} \langle l \vert H \vert j \rangle \langle j' \vert l \rangle \psi_j\psi_{j'}^* -\sum_{jj'}\langle l \vert j \rangle \langle j' \vert H \vert l \rangle \psi_j\psi_{j'}^*]
\end{equation}
which could be simplified as 

\begin{equation} \label{current-psi4}
\frac{d\rho_t^l}{dt}= I_{l+1 \rightarrow l} + I_{l-1 \rightarrow l}
\end{equation} 
where

\begin{equation} \label{current-psi41}
I_{l+1 \rightarrow l}= -\frac{1}{i\hbar} [\langle l \vert H \vert l+1 \rangle \psi_{l+1}\psi_{l}^* -\langle l+1 \vert H \vert l \rangle \psi_l\psi_{l+1}^*]
\end{equation}
and
\begin{equation} \label{current-psi42}
I_{l-1 \rightarrow l}= -\frac{1}{i\hbar} [\langle l \vert H \vert l-1 \rangle \psi_{l-1}\psi_{l}^* -\langle l-1 \vert H \vert l \rangle \psi_l\psi_{l-1}^*]
\end{equation}
The charge density is changing at atom site $l$ as a result of two currents: right moving electrons $I_{l+1}\rightarrow l$ and left moving electrons $I_{l-1}\rightarrow l$. The corresponding current to a Bloch state $\psi_j(t)=e^{ikj-iE(k)t/\hbar}$ are:

\begin{equation} \label{current-psi41}
I_{l+1 \rightarrow l}= -v_k
\end{equation}
and
\begin{equation} \label{current-psi42}
I_{l-1 \rightarrow l}= +v_k
\end{equation}
where $v_k = {\partial E(k)}/{\hbar\partial k} = 2\gamma sin(k) /\hbar$ is the group velocity. It is apparent that although the individual currents are non-zero proportional to the group velocity, the total current $I = I_{l+1 \rightarrow l} + I_{l-1 \rightarrow l}$ for a pure Bloch state is zero due to an exact balance between left and right going currents. It is worth to mention that to simplify the notation, a Bloch state $e^{ikj}$ is often normalized with its current flux $1/\sqrt{v_k}$ calculated from equation \ref{current-psi41} and \ref{current-psi42} to obtain a unitary current. Hence I will mostly use a normalized Bloch state $e^{ikj}/\sqrt{v_k}$ in later derivations.

\section{Transport on resonance and off resonance} \label{Transport on resonance and off resonance}

Nanoscale transport can be described by three regimes: 

(1) The self-consistent field (SCF) regime in which the thermal broadening $k_BT$ and coupling $\Gamma$ to the electrodes are comparable to the Coulomb energy $U_0$. The SCF method (single electron picture) implemented with NEGF could be used to describe transport in this regime as discussed in sections \ref{Breit–Wigner formula} to \ref{Equilibrium vs. non-equilibrium I-V}. In molecular junctions smaller than $\sim 3nm$, it is shown that the transport remain elastic and phase coherent at room temperature. Therefore, it is well accepted in the mesoscopic community to use SCF models to describe the properties of the molecular junctions. Based on a single electron picture and without taking into account the Coulomb energy, this NEGF method coupled to the SCF Hamiltonian describes the properties of the system on and off resonances. Good agreement between these models and many room-temperature experiments suggest applicability of this method. A simplified Breit-Wigner formula derived from this method also could be used to model on-resonances transport through the device provided the level spacing is big compared with the resonances width. However, in those cases where the Coulomb energy has higher contribution, this method cannot describe the properties of the system on resonance.

\begin{figure}[htbp] 
   \centerline{\includegraphics[width=0.8\textwidth]{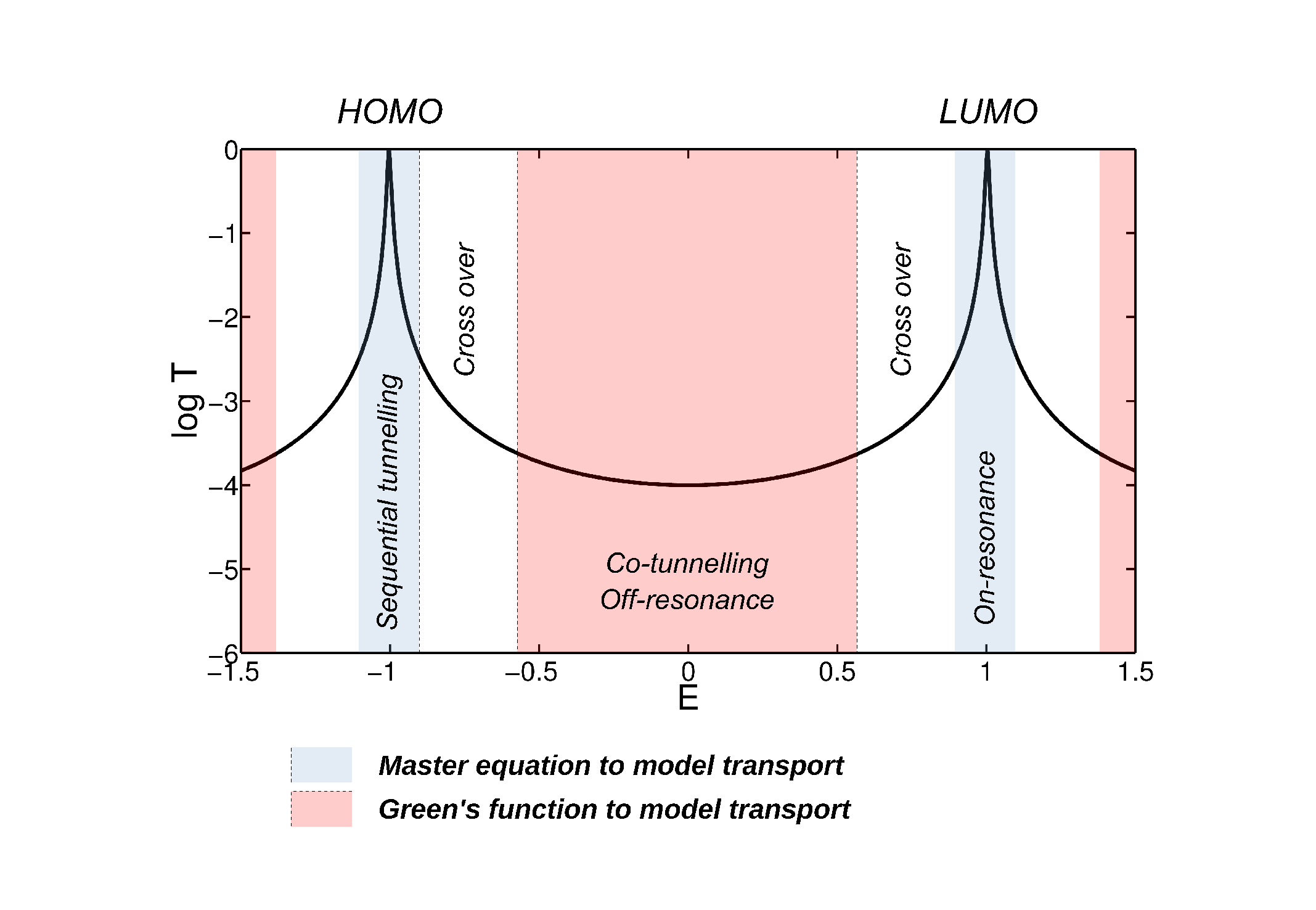}}
  \caption{\textbf{Transport on resonance and off resonance}. The transport mechanism in a molecular junction could be either in tunnelling regime (off-resonance) where electrons tunnelled through the molecule modelled usually with NEGF, or on resonance where electrons are transmitted with high rate through a energy level modelled using master equation. The intermediate state (cross-over) between on and off resonance regimes are difficult to interpret either with NEGF or master equation. }
  \label{tunneling}
\end{figure}

(2) The Coulomb blockade (CB) regime in which Coulomb energy $U_0$ is much higher than both the thermal broadening $k_BT$ and coupling $\Gamma$ where the SCF method is not adequate and the multi-electron master equation should be used to describe the properties of the system in this regime as discussed in section \ref{Master equation}. This is needed usually to model the properties of molecular junctions at low temperature where an electrostatic gate voltage could be applied through back gate.

(3) The intermediate regime in which the Coulomb energy $U_0$ is comparable to the larger of the thermal broadening $k_BT$ and coupling $\Gamma$. There is no simple approach to model this regime. Neither the SCF method nor master equation could be used to well describe the transport in this regime because SCF method does not do justice to the charging, while the master equation does not do justice to the broadening.

\subsection{Breit-Wigner formula (BWF)} \label{Breit–Wigner formula}
In the SCF regime, provided the coupling to electrodes was weak enough where the level broadening on resonances due to the electrodes are small enough and the level spacing (differences between the eigenenergies of a quantum system) is large enough, the on resonance transmission coefficient $T$ of the electrons with energy $E$ through a molecule could be described by a Lorentzian function, via the Breit-Wigner formula \cite{C4CS00203B}:

\begin{equation} \label{BWF}
T(E) = \frac{4\Gamma_1\Gamma_2}{(E-\varepsilon_n)^2+(\Gamma_1+\Gamma_2)^2}
\end{equation}
where $\Gamma_1$ and $\Gamma_2$ describe the coupling of the molecular orbital to the electrodes and $\varepsilon_n = E_n-\sigma$ is the eigenenergy
$E_n$ of the molecular orbital shifted slightly by an amount $\sigma$ due to the coupling of the orbital to the electrodes. This formula shows that when the electron resonates with the molecular orbital (e.g. when $E = \varepsilon_n$), electron transmission is a maximum. The formula is valid when the energy $E$ of the electron is close to an eigenenergy $E_n$ of the isolated molecule, and if the level spacing of the isolated molecule is larger than ($\Gamma_1+\Gamma_2$). If $\Gamma_1=\Gamma_2$ (a symmetric molecule attached symmetrically to identical leads), $T(E) = 1$ on resonance ($E = \varepsilon_n$).

If a bound state (e.g. a pendant group $\varepsilon_p$) is coupled (by coupling integral $\alpha$) to a continuum of states, Fano resonances could occur. This could be modelled by considering $\varepsilon_n=\varepsilon_0+\alpha^2/(E-\varepsilon_p)$ in BWF. At $E=\varepsilon_p$, the electron transmission is destroyed (the electron anti-resonates with the pendant orbital) and at $E=\varepsilon_n$, the electron transmission is resonated by $\varepsilon_n$. The level spacing between this resonance and antiresonance is proportional to $\alpha$. 

\subsection{Scattering theory and non-equilibrium Green's function} \label{Non-equilibrium Green's function}
Non-equilibrium Green's function method has been widely used in the literature to model electron and phonon transport in nano and molecular scale devices and has been successful to predict and explain different physical properties. The Green's function is a wave function in a specific point of the system due to an impulse source in another point. In other words, the Green's function is the impulse response of the Schrödinger equation. Therefore, a Green's function should naturally carry all information about wave-function evolution from one point to another in a system. In this paper, I have used the standard Green's function methods to calculate the transport. I will discuss it briefly but more detail discussion could be found in \cite{Claughton1995,Sanvito1999a,ferrer2014gollum,Datta2005b}.
\begin{figure}[htbp] 
   \centerline{\includegraphics[width=0.8\textwidth]{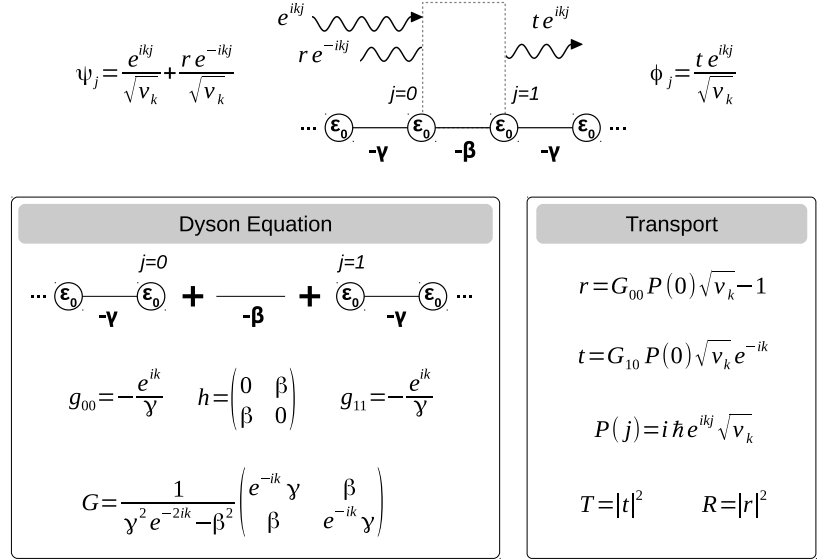}}
  \caption{\textbf{Transport through a scatter connected to two 1D leads}. For a Bloch wave $e^{ikj}/\sqrt{v_k}$ insident with a barrier, the wave is transmitted with the amplitude of $t$ ($te^{ikj}/\sqrt{v_k}$) and reflected with the amplitude of $r$ ($re^{-ikj}/\sqrt{v_k}$). Using the surface Green's function of the leads ($g_{00}$ and $g_{11}$), the Hamiltonian of the scattering region in witch bridge two leads $h$ and Dyson's equation, the total Green's function $G$ could be calculated. The Green's function is the impulse response of the system and could be used to calculate the transmission $t$ and reflection $r$ amplitudes.}
  \label{box4}
\end{figure}

Figure \ref{box4} shows how the Green's function could be used to calculate the transmission and reflection amplitudes in a two terminal system where two semi-infinite crystalline 1D leads are connected to a scattering region. The main question is what are the amplitudes of the transmitted and reflected waves? There are two main steps, first to calculate the total Green's function matrix element between the site $0$ and $1$ ($G_{10}$) or $0$ and $0$ ($G_{00}$); and secondly project these to the wavefunction to calculate transmission $t$ and reflection $r$ amplitudes. The total transmission and reflection probabilities then could be calculated by 

\begin{equation} \label{St}
T = \sum_{ij}t_{ij} t_{ij}^* = Tr (tt^\dagger)
\end{equation}
and
\begin{equation} \label{Sr}
R = \sum_{ij}r_{ij} r_{ij}^* = Tr (rr^\dagger)
\end{equation}
$t_{i,j}$ ($r_{i,j}$) is the transmission (reflection) amplitude describing scattering from the $j$th channel of the left lead to the $i$th channel of the right (same) lead. Scattering matrix $S$ is defined from $\psi_{OUT} = S \psi_{IN}$ and could be written by combining reflection and transmission amplitudes as:

\begin{equation}
S=\begin{pmatrix}r&t'\\t&r'\end{pmatrix}
\end{equation}
The $S$ matrix is a central object of \textit{scattering theory} and charge conservation implies that the $S$ matrix to be unitary: $SS^{\dagger}=I$. 

As shown in figure \ref{box4}, the total Green's function (first step) could be obtained using Dyson equation $G = (g^{-1}-h)^{-1}$ where the surface Green's functions of decoupled two semi infinite leads $g=\bigl(\begin{smallmatrix} g_{00} & 0 \\ 0 & g_{11} \end{smallmatrix}\bigr)$ and the Hamiltonian in which couples them together $h$ are known. The second step is to calculate the projector which projects the Green's function in the leads 

\begin{equation} \label{1Dlead-GF}
g = \sum_{jl} g_{jl}\vert j \rangle \langle l \vert = \sum_{jl} \frac{e^{ik\vert j-l \vert}}{i\hbar v_k}\vert j \rangle \langle l \vert
\end{equation}
to the normalized wavefunction at site $l$ ($e^{ikl}$). It could be shown that \cite{Claughton1995,Sanvito1999a}, this projector $P(j)$ also projects the total Green's function $G$ to the wavefunction $\psi$ and therefore could be used to calculate $t$ and $r$. Using this projector at site $j=0$, $P(0)$ and $G_{10}$ ($G_{00}$), the transmission (reflection) amplitude is obtained (see figure \ref{box4}).

%
%


\subsection{The Landauer Formula} \label{The Landaur Formula}

Landauer used the \textit{scattering theory of transport} as a conceptual framework to describe the electrical conductance and wrote "Conductance is transmission" \cite{Landauer1987}. In the Landauer approach a mesoscopic scatterer is connected to two ballistic leads (see figure \ref{box0}). The leads are connected to the reservoirs where all inelastic relaxation processes take place. The reservoirs have slightly different electrochemical potentials $\mu_{L}-\mu_{R}\rightarrow 0$ to drive electrons from the left to the right lead. The current therefore could be written as:

\begin{equation} \label{land-I}
I=\frac{e}{h} \int dE\: T(E) \left(f(E-\mu_{L})-f(E-\mu_{R})\right)
\end{equation}
where $e$ is the electronic charge, $T(E)$ is the transmission coefficient and $f$ is Fermi-Dirac distribution function $f(E-\mu)=1/(1+e^{(E-\mu)/k_BT})$ associated with the electrochemical potential $\mu$, $k_B$ is Boltzmann constant and $T$ is temperature. The Fermi functions can be Taylor expanded over the range $eV$,

\begin{equation} \label{land-I2}
I=\frac{e}{h} \int dE\: T(E) \left(-\frac{\partial f(E)}{\partial E}\right) (\mu_{L}-\mu_{R})
\end{equation}
where $\mu_{L}-\mu_{R}=eV$. By including the spin, the electrical conductance $G = I/V$ reads as:

\begin{equation} \label{land-G}
G=\frac{2e^2}{h} \int dE\: T(E) \left(-\frac{\partial f(E)}{\partial E}\right)
\end{equation}
At $T=0K$, $-\frac{\partial f(E-\mu)}{\partial E} = \delta (\mu)$ where $\delta (\mu)$ is the Kronecker delta. For an ideal periodic chain where $T(E) = 1$ at $T=0K$, the Landauer formula becomes:

\begin{equation} \label{land-G00}
G_0=\frac{2e^2}{h} \simeq 77.5 \text{ $\mu$ Siemens}
\end{equation}
$G_0$ is called the "\textit{Conductance Quantum}". In other words, the current associated with a single Bloch state $v_k/L$ and generated by the electrochemical potential gradient is $I=e(v_k/L)D\Delta\mu$ where the density of states $D = \partial n/\partial E=L/hv_k$. It is worth mentioning that the Landauer formula \ref{land-I2} describes the linear response conductance, hence it only holds for small bias voltages, $\delta V \rightarrow 0$. 

\subsubsection{Landauer-Buttiker formula for multi-terminal structuers} \label{Landauer-Buttiker formula for multi-terminal structuers}

Conductance measurements are often performed using a four-probe structure to minimize the contact resistance effect. Also multi-probe structures are widely used to describe the Hall-effect or in sensing applications. Based on the Landauer approach for two terminal system, Buttiker \cite{Buttiker1988} suggested a formula to model multi-probe currents for structures with multiple terminals as:

\begin{equation} \label{Buttiker-G}
I_i = \frac{e}{h}\sum_j T_{ij}(\mu_i-\mu_j)
\end{equation}
where $I_i$ is the current at $i$th terminal and $T_{ij}$ is the transmission probability from terminal $j$ to $i$. In a multi-terminal system, it is consistent to assume one of the probes as reference voltage $V_{ref} = 0$ and write the currents based on that. As an example, for a four probe structure, the current in each probe could by written as:

\begin{equation} \label{4pG}
\begin{pmatrix}
I_1\\ 
I_2\\ 
I_3\\ 
I_4
\end{pmatrix}
=\frac{2e^2}{h}
\begin{pmatrix}
N_1-T_{11} & -T_{12} & -T_{13} & -T_{14}\\ 
-T_{21} & N_2-T_{22} & -T_{23} & -T_{24}\\ 
-T_{31} & -T_{32} & N_3-T_{33} & -T_{34}\\ 
-T_{41} & -T_{42} & -T_{43} & N_4-T_{44}
\end{pmatrix}
\begin{pmatrix}
V_1\\ 
V_2\\ 
V_3\\ 
V_4
\end{pmatrix} 
\end{equation}
where $N_i$ is number of open conduction channels in lead $i$. In a four probe structure, if probe 3 and 4 are outer voltage probes ($I_3 = I_4 = 0$) and probe 1 and 2 are the inner current probes, the four probe conductance is $G_{four-probe} = (2e^2/h)(V_3-V_4)/I_1$. 

\subsection{Generalized model to calculate T(E)} \label{Generalized model to calculate T(E)}

In this section, I would like to discuss the generalized approach to calculate the transmission coefficient $T$ of the electrons (phonons) with energy $E$ ($\hbar \omega$) passing from one electrode to another using non-equilibrium Green's function method. Consider a quantum structure connected to ideal, normal leads of constant cross-section, labelled $L=1, 2, \dots$ and therefore begin by considering two vector spaces $A$ (representing the normal leads) and $B$ (representing the structure of interest), spanned by a countable set of basis functions. For a system with an orthogonal basis set where the overlap matrix is unitary matrix $I$, the expression for the transmission coefficient $T_{nn'}$ between two scattering channels $n, n'$ of an open vector space $A$, in contact with a closed sub-space $B$ could be written as \cite{Claughton1995}:

\begin{figure}[htbp] 
   \centerline{\includegraphics[width=1.05\textwidth]{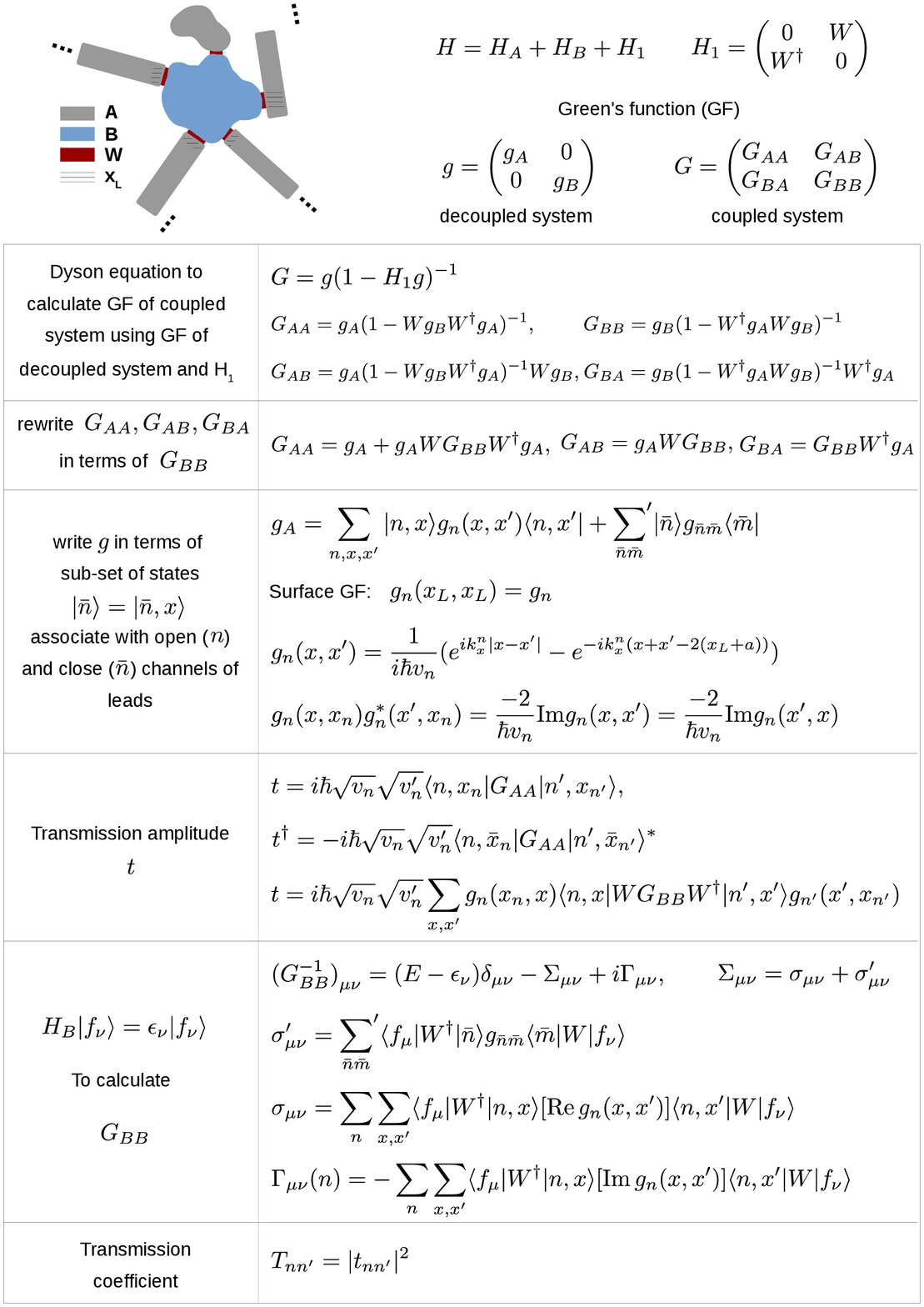}}
  \caption{\textbf{Generalized transport model using Non-equilibrium Green's function method}\cite{Claughton1995}.}
  \label{box7}
\end{figure}

\begin{equation} \label{Tfromtt0}
T_{nn'}=\vert t_{n,n'}(E,H)\vert^2
\end{equation}
As shown in figure \ref{box7}, the transmission amplitudes could be written \cite{Claughton1995} using the surface Green's function in the leads $A$ and the Green's function of the scattering region $B$ coupled to the outside world through coupling matrix elements $H_1$.

\begin{equation} \label{Tfromtt01}
t_{nn'} = i\hbar\sqrt{v_n} \sqrt{v'_n} \langle n \vert g W G_{BB} W^\dagger g \vert n' \rangle
\end{equation}
or more precisely

\begin{equation} \label{Tfromtt1}
t_{nn'} = i\hbar\sqrt{v_n} \sqrt{v'_n} \sum_{x,x^\prime}g_n(x_n,x)\langle n,x\vert { W G_{BB} W^\dagger}
\vert n',x'\rangle g_{n'}(x',x_{n'})
\end{equation}
where 

\begin{equation} \label{Tfromtt2}
{(G_{BB}^{-1})}_{\mu\nu}=(E-\epsilon_\nu)\delta_{\mu\nu}-\Sigma_{\mu\nu}+i\Gamma_{\mu\nu}, 
\end{equation}
and 

\begin{equation} \label{Tfromtt3}
g_{n}(x,x')=\frac{e^{i k_x^{n}\vert x-x'\vert}
-e^{-i k_x^{n}( x+x'-2(x_L+a))}}{i\hbar v_{n}}
\end{equation}
is the Green's function of the semi-infinite lead between any position point $x$ and $x'$ in the transport direction terminated at $x = x_L$ and vanishes at $x = x_L+a$. $k_x^{n}$ is the longitudinal wavevector of channel $n$. If the lead belonging to channel $n$ terminates at $x = x_L$, then on the surface of the lead, the Green's function $g_n(x,x')$ takes the form $g_n(x_L,x_L)=g_n$, where $g_n = a_n+ib_n$ with $a_n$ real and $b_n$ equal to $\pi$ times the density of states per unit length of channel $n$. Moreover, if $v_n$ is the group velocity for a wavepacket travelling along channel $n$, then $\hbar v_n = 2b_n/{\vert g_n \vert}^2$. It is interesting to note that if $x$ and $x'$ are positions located between $x_L$ and some point $x_n$,

\begin{equation} \label{Tfromtt31}
g_n(x,x_n)g^*_n(x^\prime,x_n)={-2\over\hbar v_n} {\rm Im} g_n(x,x^\prime)
={-2\over\hbar v_n} {\rm Im} g_n(x^\prime,x)
\end{equation}
The eigenvalue and eigenvectors associated with the Hamiltonian of the $B$ is obtained from the Schr\"{o}dinger equation $H_B\vert f_\nu\rangle=\epsilon_\nu\vert f_\nu\rangle$. The self-energies $\Sigma$ and broadening $\Gamma$ then could be written as \cite{Claughton1995}:

\begin{equation} \label{Tfromtt4}
\Sigma_{\mu\nu}= 
\sum_n \sum_{x,x^\prime}\langle f_\mu\vert
{ W^\dagger}
\vert n,x\rangle [{\rm Re\,}g_n(x,x^\prime)]
 \langle n,x^\prime\vert { W}\vert f_\nu \rangle
 +
{\sum_{{\bar n}{\bar m}}}^\prime
\langle f_\mu\vert
{ W^\dagger}\vert{\bar n}\rangle g_{{\bar n}{\bar m}}
 \langle{\bar m}\vert { W}\vert f_\nu \rangle 
\end{equation}
and

\begin{equation} \label{Tfromtt5}
 \Gamma_{\mu\nu}(n)=
-\sum_n\sum_{x,x^\prime}\langle f_\mu\vert
{ W^\dagger}
\vert n,x\rangle [{\rm Im\,}g_n(x,x^\prime)]
 \langle n,x^\prime\vert { W}\vert f_\nu \rangle
\end{equation}
\begin{figure}[htbp] 
   \centerline{\includegraphics[width=1.1\textwidth]{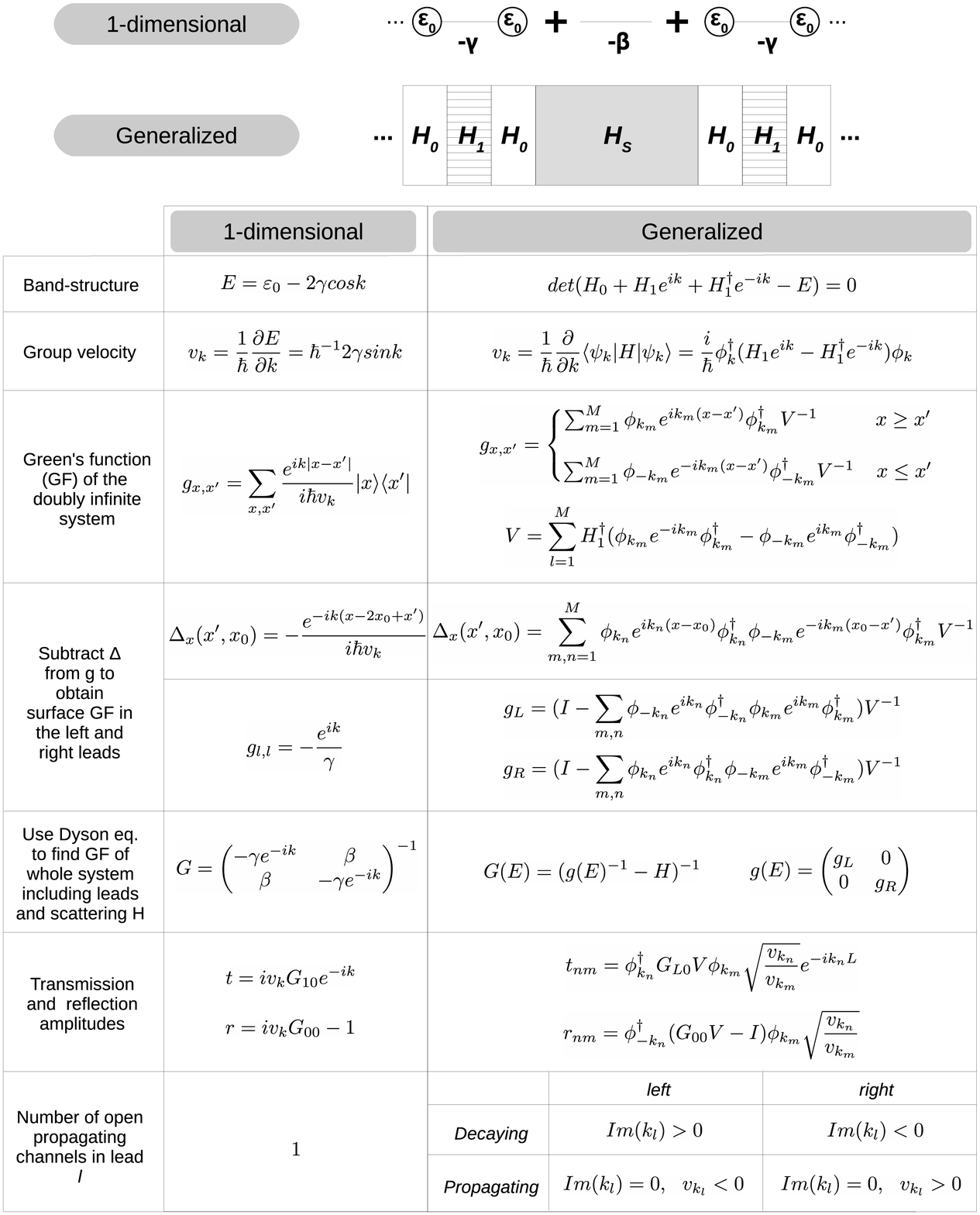}}
  \caption{\textbf{Generalized transport model using Green's function method}. Generalized transport model using equilibrium Green's function method \cite{Sanvito1999a} and its equivalent model for a simple 1D problem.}
  \label{box5}
\end{figure}
This is very general and makes no assumptions about the presence or otherwise of resonances. For a system with non-orthogonal basis states, in equation \ref{Tfromtt2} $\delta_{\mu\nu}$ should be replaced with the overlap matrix $S_{\mu\nu} = \langle f_\mu \vert f_\nu\rangle$. It is interesting to note that the vector spaces A representing the normal leads include both crystalline structures connected to the outside world and any close system coupled to the vector spaces B representing the structure of interest. In the latter case, the only effect of the closed part of the vector spaces A is to contribute in the scattering by its self-energy. The physical meaning of this and where it could be useful are discussed more in the next section. Furthermore, figure \ref{box5} shows a slightly different approach to calculate the transmission (reflection) amplitude $t$ ($r$) in a two terminal system with non-orthogonal basis set derived in \cite{Sanvito1999a}. For better understanding, as well as the most general approach a simplified specific case for a one dimensional lead connected to an arbitrary scattering region is also included in this figure.


\subsection{Equilibrium vs. non-equilibrium I-V} \label{Equilibrium vs. non-equilibrium I-V}
There are the terms usually used in the literature such as elastic vs. inelastic processes, coherent vs. incoherent regime or equilibrium vs. non-equilibrium Green's function method. The average distance that an electron (or a hole) travels before changing its momentum (energy) called elastic (inelastic) mean free path. For a junction with the length smaller than elastic (or inelastic) mean free path the process is assumed to be ballistic. These definition are well accepted in the mesoscopic community. However, the equilibrium and non-equilibrium process are defined differently in the literature. The view I adopt is to call any process where the current is derived from any differences in the electrochemical potential whether small or big is called non-equilibrium condition. To calculate the current using Landauer formula (equation \ref{land-I}), one needs to bear in mind that the Landauer formula only holds in the linear response regime for a transmission coefficient $T$ which describes the transmission probability of particle with energy $E$ from one electrode to another calculated in steady state condition and assuming the junction is close to equilibrium ($\delta V \rightarrow 0$). However, for the non-linear regime where the voltage condition is big, the transmission coefficient $T$ could be a function of bias voltages $V_b$. The potential profile applied to the junction due to a given electric field caused by bias voltage should be calculated by Poisson's equation \cite{Datta2005b}. In the non-equilibrium condition, the Landauer formula then takes the form,

\begin{equation} \label{land-IVb}
I(V_b,V_g)=\frac{e}{h} \int dE\: T(E,V_b,V_g) \left(f(E+\frac{eV_b}{2})-f(E-\frac{eV_b}{2})\right)
\end{equation}
It is worth mentioning that in some experiments due to very noisy measured conductance spectrum $G = I/V_b$, the differential conductance map $G_{diff}(V_b,V_g) = dI(V_b,V_g)/dV_b$ is plotted which could be calculated by differentiation of equation \ref{land-IVb} with respect to the bias voltage $V_b$. 

\begin{figure}[htbp] 
   \centerline{\includegraphics[width=0.7\textwidth]{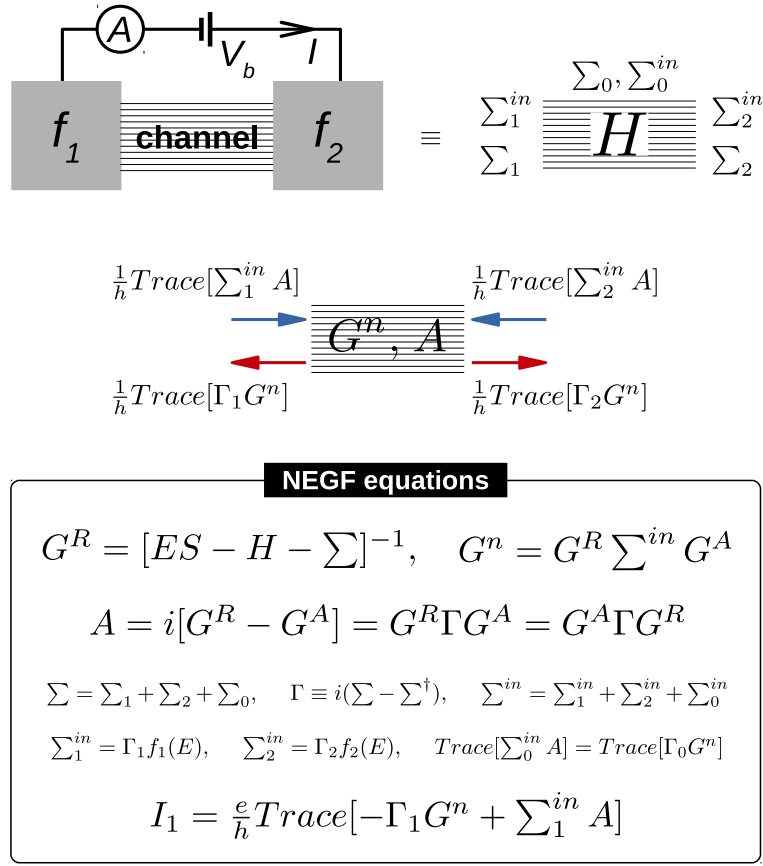}}
  \caption{\textbf{Non-equilibrium Green's function (NEGF) equations}.}
  \label{box10}
\end{figure}

Another interesting point is how to interpret transport in a nano and molecular scale junctions physically. If $ES\vert\psi\rangle = H\vert\psi\rangle$ describes the properties of the closed system $H$ with non-orthogonal basis set $S$, then once it connects to the outside world and became an open system (see figure \ref{box10}), the modified Schr\"{o}dinger equation in non-equilibrium condition could be written \cite{Datta2005b}:

\begin{equation} \label{schmod}
ES\vert\psi\rangle = H\vert\psi\rangle+\Sigma\vert\psi\rangle+\vert s \rangle
\end{equation}
where the terms $\Sigma\vert\psi\rangle$ and $\vert s \rangle$ describe the outflow and inflow, respectively arises from the boundary conditions. Equation \ref{schmod} could be rewritten as 

\begin{equation} \label{schmod0}
\vert\psi\rangle = [G^R]\vert s \rangle
\end{equation}
where $G^R=[ES-H-\Sigma]^{-1}$ is retarded Green's function ($G^A = [G^R]^\dagger$), $\Sigma = \Sigma_1+\Sigma_2+\Sigma_0$ is self-energies due to the electrodes $\Sigma_1$, $\Sigma_2,$ and surroundings $\Sigma_0$ such as dephasing contact or inelastic scattering e.g. electron-phonon coupling, emission, absorption, etc. Dephasing contact terms could be described by SCF method whereas for inelastic processes one needs to use for instance Fermi's golden rule to describe these self energies. 
There are some disagreement in the literature about how to treat incoherent and inelastic processes \cite{Datta2005b,Buttiker1988a}. Buttiker's \cite{Buttiker1988a} view is to treat the inelastic and incoherent scattering by introducing a new electrode to the original coherent system. This could be seen as assigning the new self-energies associated with any inelastic or incoherent process. However, Datta has slightly different view. If you treat the incoherent and inelastic effect by introducing an extra electrode, you assign a corresponding distribution function e.g. Fermi function for electrons which in general may not be the case. More generally, you could introduce any incoherence and/or inelastic process by appropriate self energy which not necessarily  described by equivalent Fermi function in the contact. 

For a normal, coherent elastic junction if $H_{1,2}$ are the coupling matrices between electrode 1 (2) and scattering region and $g_{1,2}$ are the surface Green's function of the electrodes, $\Sigma_{1,2} = H^\dagger_{1,2} g_{1,2} H_{1,2}$. Furthermore, the current could be calculated as $I_1 = \frac{e}{h} Trace[-\Gamma_1G^n+\sum_1^{in}A]$ where $\Gamma_1$, $G^n$, $\sum_1^{in}$ and $A$ defined in figure \ref{box10}. From the basic law of equilibrium, in a special situation where we have only one contact connected; the ratio of the number of electrons to the number of states must be equal to the Fermi function in the contact ($\sum_{1,2}^{in} = \Gamma_{1,2}f_{1,2}(E)$). However, in dephasing contact, $\Sigma_0^{in}$ is not described by any Fermi function and since inflow and outflow should be equal $Trace[\sum_0^{in}A]=Trace[\Gamma_0G^n]$. Figure \ref{box10} summarize the basic non-equilibrium Green's function (NEGF) equations to calculate the current in a most general junction where surroundings presents. In the absence of surroundings, current in lead $i$ could be written as \cite{Datta2005b}:

\begin{equation} \label{schmod1}
I_i = \frac{e}{h} \sum_j Trace[\Gamma_i G^R \Gamma_j G^A](f_i-f_j)
\end{equation}
where $T_{ij}(E)=Trace[\Gamma_i(E) G^R(E) \Gamma_j(E) G^A(E)]$ is the transmission coefficient for electrons with energy $E$ passing from lead $i$ to lead $j$.
\begin{figure}[htbp] 
   \centerline{\includegraphics[width=0.8\textwidth]{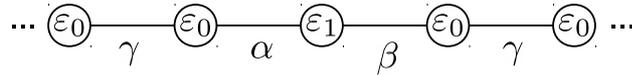}}
  \caption{\textbf{Two terminal system with two 1D leads connected to a scattering region $\varepsilon_1$.}}
  \label{box12}
\end{figure}
Consider two identical 1D leads with on-site energies $\varepsilon_0$ and hoping integrals $\gamma$ connected to a scattering region $\varepsilon_1$ with coupling integrals $\alpha$ and $\beta$ as shown in figure \ref{box12}. The transmission coefficient $T$ for electrons with energy $E$ traversing from left to right lead can be calculated as 

\begin{equation} \label{schmod11}
T(E)=\Gamma_L(E) G^R(E) \Gamma_R(E) G^A(E)
\end{equation}
where the retarded Green's function is $G^R(E) = (E-\varepsilon_1-\Sigma)$, the self-energies $\Sigma = \Sigma_L+\Sigma_R$ obtained from $\Sigma_L = \alpha^2 e^{ik}/\gamma$ and $\Sigma_R = \beta^2 e^{ik}/\gamma$ and the broadening due to the left and right leads are $\Gamma_L = i(\Sigma_L-\Sigma_L^\dagger)=-2\alpha^2sin(k)/\gamma$ and $\Gamma_R = i(\Sigma_R-\Sigma_R^\dagger)=-2\beta^2sin(k)/\gamma$. 

\subsection{Master equation} \label{Master equation}
In the multi-electron picture, the overall system has different probabilities $P_\alpha$ of being in one of the $2^N$ possible states $\alpha$. Furthermore all of these probabilities $P_\alpha$ must add up to one. The individual probabilities could be calculated under steady-state conditions where there is no net flow into or out of any state (see figures \ref{box8} and \ref{box9})

\begin{equation} \label{master_eq}
\sum_\beta R(\alpha \rightarrow \beta) P_\alpha = \sum_\beta R(\beta \rightarrow \alpha) P_\beta
\end{equation}
where $R(\alpha \rightarrow \beta)$ is the rate constants obtained by assuming a specific model for the interaction with the surroundings. In a system that the electrons can only enter or exit from the source and drain contacts, these rates are given in figures \ref{box8} and \ref{box9} for one and two level systems. This equation is called a multi-electron master equation \cite{Datta2005b}. 

\subsubsection{One level system} \label{One level system}
One-electron energy levels represent differences between energy levels corresponding to states that differ by one electron. If $E(N)$ is the energy associated with the $N$-electron state, the energy associated with the addition (removal) of one electron are called affinity (ionization) energy. 

\begin{equation} \label{affinity-ionization}
\begin{aligned}
IP = E(N-1)-E(N),\\ 
EA = E(N)-E(N+1)
\end{aligned}
\end{equation}
The energy-gap $E_g$ of a molecule (sometimes called additional energy) could be calculated from $IP$ and $EA$ as: $E_g = IP - EA$ \cite{Datta2005b}. The important conceptual point is that the electrochemical potential $\mu$ should lie between the affinity levels (above $\mu$) and ionization levels (below $\mu$). Figure \ref{box8} shows the master equation for spin-degenerate one level system with energy $\varepsilon$ where there are only two possibilities, either the state is full $\vert 1 \rangle$ or empty $\vert 0 \rangle$. The current then could be calculated as:
\begin{figure}[htbp] 
   \centerline{\includegraphics[width=0.55\textwidth]{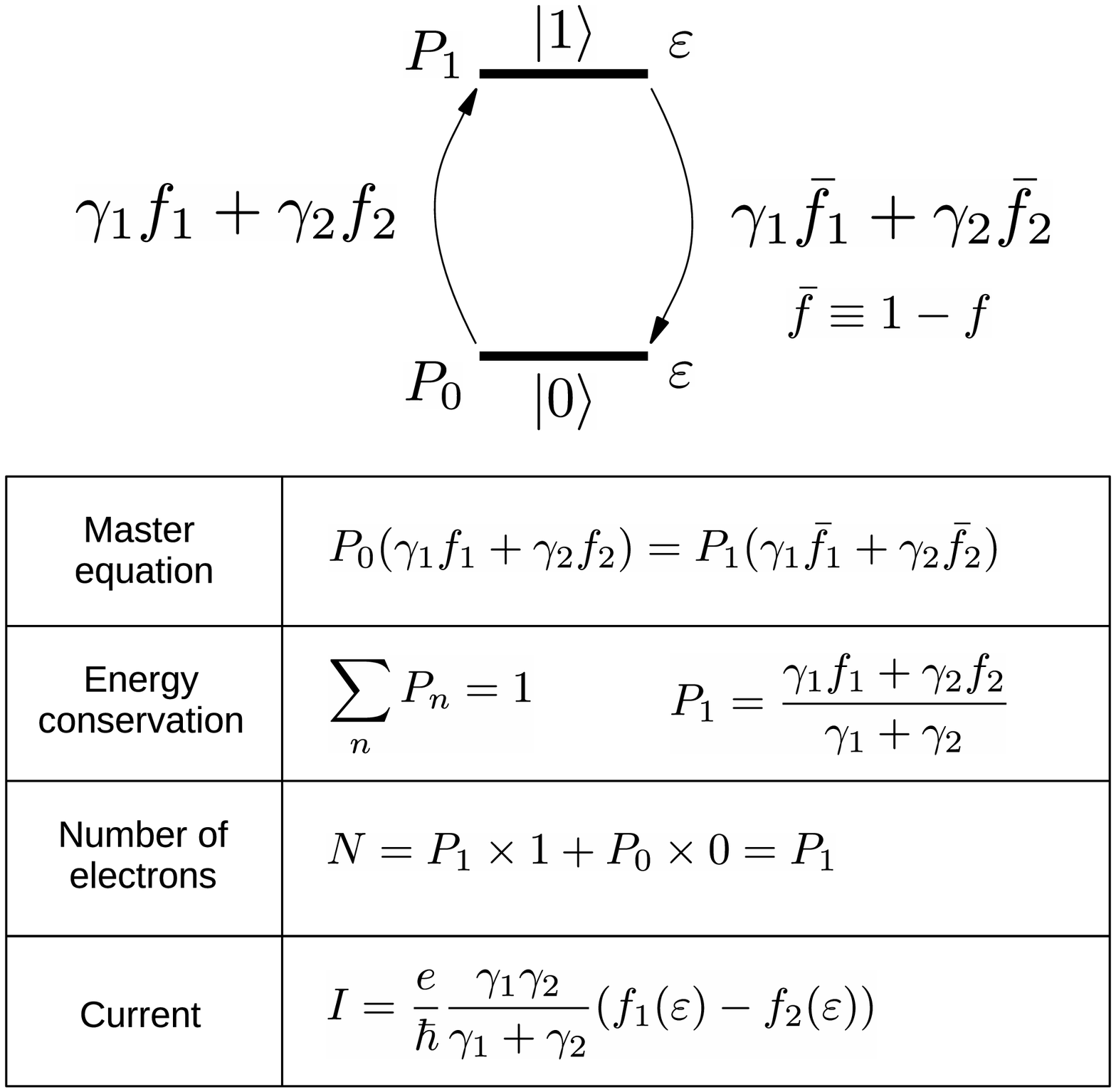}}
  \caption{\textbf{One level system}.}
  \label{box8}
\end{figure}

\begin{equation} \label{master1level}
I=\frac{e}{\hbar}\frac{\gamma_1\gamma_2}{\gamma_1 + \gamma_2} (f_1(E)-f_2(E))
\end{equation}
where $\gamma_1$ and $\gamma_2$ are the rates electron can go in and out from the left and right electrodes with $f_1(E)$ and $f_2(E)$ Fermi functions.

\subsubsection{Two level system} \label{Two level system}
\begin{figure}[htbp] 
   \centerline{\includegraphics[width=0.65\textwidth]{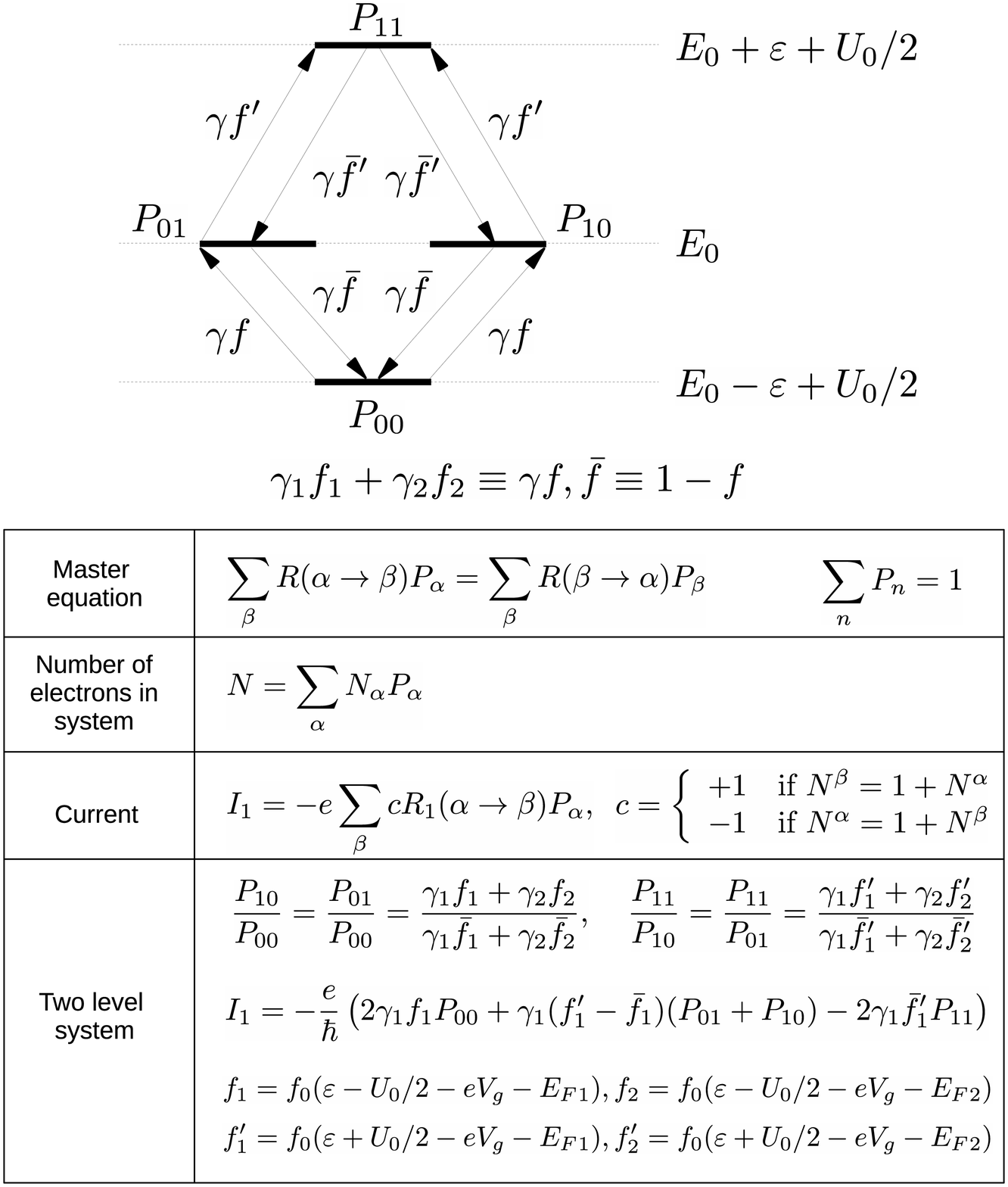}}
  \caption{\textbf{Two level system}.}
  \label{box9}
\end{figure}
However, in two level system there are four possibilities, both empty $\vert 00 \rangle$ or full $\vert 11 \rangle$ and either one of them full and another empty ($\vert 01 \rangle$ and $\vert 10 \rangle$). Figure \ref{box9} shows the obtained current for two level system \cite{Datta2005b}. The crucial point here is that, as soon as one state is full, there need an additional energy (Coulomb repulsion energy) to have second electron in the another state in addition to the level spacing energy. Another conceptual point is, it is incorrect to assume one Fermi function for all transitions. Due to the Coulomb blockade energy, each level needs certain electrochemical potential to overcome the barrier and current flow. 

\subsubsection{Coulomb and Franck-Condon blockade regimes}
The electronic properties of weakly coupled molecules are dominated by Coulomb interactions and spatial confinement at low temperatures. This could lead to Coulomb blockade (CB) regimes in which the channel is blocked due to the presence of an electron trapped in the channel. In addition, charge transfer can excite vibrational modes or vibrons, and strong electron-vibron coupling leads to suppression of tunnel current at low bias called Franck-Condon (FC) blockade regimes. 

To describe the transport in this regime, a minimal model (the Anderson-Holstein Hamiltonian) could be used \cite{Koch2005a} that captures the CB, FC and the Kondo effect if three assumptions are made: (1) the relaxation in the leads  assumed to be sufficiently fast leading to Fermi functions for the distribution of the electrons in thermal equilibrium at all times; (2) the transport through the molecule is dominated by tunneling through a single, spin-degenerate electronic level, and (3) one vibron taken into account within the harmonic approximation. In this case, the Anderson-Holstein Hamiltonian reads $H = H_{mol} + H_{leads} + H_T$ with

\begin{equation} \label{master-CB-FC0}
H_{mol} = \varepsilon_d n_d + U n_{d\uparrow}n_{d\downarrow}+\hbar\omega b^\dagger b + \lambda \hbar \omega (b^\dagger +b)n_d
\end{equation}
describing the electronic and vibrational degrees of freedom of the molecule,

\begin{equation} \label{master-CB-FC1}
H_{leads} = \sum_{a=L,R} \sum_{p,\sigma} (\varepsilon_{ap}-\mu_a)c_{ap\sigma}^\dagger c_{ap\sigma}
\end{equation}
the noninteracting leads, and

\begin{equation} \label{master-CB-FC2}
H_{T} = \sum_{a=L,R} \sum_{p,\sigma} (t_{ap}c_{ap\sigma}^\dagger d_\sigma + h.c.)
\end{equation}
the tunneling between the leads and molecule. Here, Coulomb blockade is taken into account via the charging energy $U$ where $eV,k_BT<<U$. The operator $d_\sigma$ ($d_\sigma^\dagger$) annihilates (creates) an electron with spin projection $\sigma$ on the molecule, $n_d = \sum_\sigma d_\sigma d_\sigma^\dagger$ denotes the corresponding occupation-number operator. Similarly, $c_{ap\sigma}$ ($c_{ap\sigma}^\dagger$) annihilates (creates) an electron in lead $a$ ($a=L,R$) with momentum $p$ and spin projection $\sigma$. Vibrational excitations are annihilated (created) by $b$ ($b^\dagger$). They couple to the electric charge on the molecule by the term $\sim n_d (b^\dagger + b)$, which can be eliminated by a canonical transformation, leading to a renormalization of the parameters $\varepsilon$ and $U$, and of the lead-molecule coupling $t_a \rightarrow t_a e^{-\lambda (b^\dagger + b)}$. The master equations determining the molecular occupation probabilities $P_q^n$ for charge state $n$ and vibrons $q$ is:

\begin{equation} \label{master-CB-FC3}
\frac{dP_q^n}{dt}=\sum_{n',q'}(P_{q'}^{n'}W_{q'\rightarrow q}^{n' \rightarrow n}-P_{q}^{n}W_{q\rightarrow q'}^{n \rightarrow n'}) -\frac{1}{\tau} (P_q^n-P_q^{eq}\sum_{q'}P_{q'}^n)
\end{equation}
$P_q^{eq}$ denotes the equilibrium vibron distribution with a relaxation time $\tau$ and $W_{q\rightarrow q'}^{n \rightarrow n'}$ denotes the total rate for a transition from $\vert n,q \rangle$ to $\vert n',q' \rangle$.

\begin{equation} \label{master-CB-FC4}
\begin{aligned}
W_{q\rightarrow q'}^{n \rightarrow n+1} = \sum_{a=L,R}(f_a(E_{q'}^{n+1}-E_q^n))\Gamma_{q\rightarrow q';a}^{n \rightarrow n+1},  \\
W_{q\rightarrow q'}^{n \rightarrow n-1} = \sum_{a=L,R}(1-f_a(E_{q}^{n}-E_{q'}^{n-1}))\Gamma_{q\rightarrow q';a}^{n \rightarrow n-1}
\end{aligned}
\end{equation}
where $f_a$ is the Fermi function and the transition rates $\Gamma$ are calculated from Fermi's golden rule.

\begin{equation} \label{master-CB-FC5}
\begin{aligned}
\Gamma_{q\rightarrow q';a}^{n \rightarrow n+1} = s^{n \rightarrow n+1} \frac{2\pi}{\hbar}\rho_a(E_{q'}^{n+1}-E_q^n) |M_{q\rightarrow q';a}^{n \rightarrow n+1}| \\
\Gamma_{q\rightarrow q';a}^{n \rightarrow n-1} = s^{n \rightarrow n-1} \frac{2\pi}{\hbar}\rho_a(E_{q}^{n}-E_{q'}^{n-1}) |M_{q\rightarrow q';a}^{n \rightarrow n-1}| 
\end{aligned}
\end{equation}
Here, $\rho_a$ denotes the density of states in lead $a$,  $M_{q\rightarrow q';a}^{n \rightarrow n\pm 1}$  denotes the FC matrix elements and $s^{n\rightarrow m}$ the spin factor \cite {lau2015redox} such that for sequential tunnelling and assuming twofold degeneracy they are $s^{1\rightarrow 0} = s^{1\rightarrow 2} =1, s^{0\rightarrow 1}= s^{2\rightarrow 1}=2$. The matrix elements $M_{q\rightarrow q';a}^{n \rightarrow n\pm 1}$ defined for vibrations are

\begin{equation} \label{master-CB-FC6}
\begin{aligned}
M_{q\rightarrow q';a}^{n \rightarrow n\pm 1}=t_0\sqrt{\frac{q_1!}{q_2!}}\lambda^{q_2-q_1} e^{-\lambda^2/2}
\end{aligned}
\end{equation}
where $q_1=min\{q,q'\}$ and $q_2=max\{q,q'\}$.

\section{Modelling the experiment} \label{Modelling the experiment}
So far I have briefly discussed, different  transport regimes and the methods to model electron and phonon through nanoscale junctions.  However, all these tools are only useful if they can explain new physical phenomenon or predict a new characteristic for a future physical system. Experiments in the field of molecular electronics either study new junction  physical properties such as conductance and current or they focus on using well characterized junctions for future applications. The crucial point is, there are certain phenomenon that only theory could access and analyse such as wave-functions, which is not a physical  observable  and others that only experiment could shed light, such as the position  of the Fermi energy, the overall effect of the inhomogeneous broadening on the transport, or screening effects which is related to the exact junction configuration in the real-time experiment. Therefore, theoretically, predictions  made based for the trends by comparing two or more system with the similar condition are potentially  more reliable than those which are only based on the numbers predicted from  the theory. 

The bottom line is the theory and experiment are not two isolated endeavours. They need to talk to each-other to lead a new discoveries. Those quantities that cannot be  computed reliably, but for which experimental  data is available, can be used to correct and refine theoretical models. Usually to explain  new phenomena, one needs to make a working  hypothesis and then try to build a model to quantify the phenomenon. To make an initial hypothesis, a theorist needs to know how different physical phenomenon such as the effect of the environment,  presence of an electric or magnetic field could be modelled. In the following, my aim is to make a few bridges between the well-known  physical phenomena and the methods to model them theoretically.

\subsection{Virtual leads versus physical leads}
Let's start by considering the differences between a lead and a channel theoretically? From a mathematical viewpoint, channels connect an extended scattering region to a reservoir and the role of lead $i$ is simply to label those channels $k_i, \bar q_i$, which connect to a particular reservoir $i$. Conceptually, this means that from the point of view of solving a scattering problem at energy $E$, a single lead with $N(E)$ incoming channels can be regarded as $N(E)$ virtual leads, each with a single channel. We could take advantage  of this equivalence by regarding the above groups of channels with wave-vectors $k_{\alpha_i}, \bar q_{\alpha_i}$ as virtual  leads and treating them on the same footing as physical leads. 

This viewpoint is particularly useful when the Hamiltonians $H_0^i$, $H_1^i$ describing the principle layers PLs (the identical periodic unit cells $H_0^i$ connected to each other by $H_1^i$) of the physical lead $i$ are block diagonal with respect to the quantum numbers associated with $k_{\alpha_i}, \bar q_{\alpha_i}$. For example, this occurs when the leads possess a uniform magnetization, in which case the lead Hamiltonian is block diagonal with respect to the local magnetization axis of the lead and $\alpha$ represents the spin degree of freedom $\sigma$. This occurs also when the leads are normal metals, but the scattering region contains one or more superconductors, in which case the lead Hamiltonian is block diagonal with respect to particle and hole degrees of freedom and $\alpha$ represents either particles $p$ or holes $h$. More generally, in the presence of both magnetism and superconductivity, $\alpha$ would represent combinations of spin and particles and holes degrees of freedom.

In all of these cases, $H_0^i$, $H_1^i$ are block diagonal and it is convenient to identify virtual leads $\alpha_i$ with each block, because I can compute the channels $k_{\alpha_i}, \bar q_{\alpha_i}$ belonging to each block in separate calculations and therefore guarantees that all such channels can be separately identified. This is advantageous, because if all channels of $H_0^i$, $H_1^i$ were calculated simultaneously, then in the case of degeneracies, arbitrary superpositions of channels with different quantum numbers could result and therefore it would be necessary to implement a separate unitary transformation to sort channels into the chosen quantum numbers. By treating each block as a virtual lead, this problem is avoided. 

\subsection{Charge, spin and and thermal currents}
When comparing theory with experiment, we are usually interested in computing the flux of some quantity $Q$ from a particular reservoir. If the amount of $Q$ carried by quasi-particles of type $\alpha_i$ is $Q_{\alpha_i}(E)$, then the flux of $Q$ from reservoir $i$ is:

\begin{equation}
I^{i}_{Q}=\int(dE/h) \sum_{\alpha_i, j,\beta_j}P^{i,j}_{\alpha_i,\beta_j} \bar f^j_{\beta_j}(E)  \label{xxxx22}
\end{equation}
$P^{i,j}_{\alpha_i,\beta_j}$ in this expression is transmission coefficient of quasi-particles of type $\alpha_i$. In the simplest case of a normal conductor, choosing $Q_{\alpha_i} = -e$ , independent of $\alpha_i$, this equation yields the electrical current from lead $i$. $\alpha_i$ may represent spin, and in the presence of superconductivity it may represent hole ($\alpha_i = h$) or particle ($\alpha_i = p$) degrees of freedom. In the latter case, the charge $Q_p$ carried by particles is $-e$, whereas the charge $Q_h$ carried by holes is $+e$. In the presence of non-collinear magnetic moments, provided the lead Hamiltonians are block diagonal in spin indices,
choosing $\alpha_i =\sigma_i$ and $Q_{\alpha_i}=-e$ in Eq. (\ref{xxxx22}) yields for the total electrical current

\begin{equation}
I^{i}_{e}=-e\int(dE/h) \sum_{\sigma_i, j,\sigma_j}P^{i,j}_{\sigma_i,\sigma_j} \bar f^j_{\beta_j}(E)  \label{xxxx1}
\end{equation}
Note that in general it is necessary to retain the subscripts $i,j$ associated with $\sigma_i$ or $\sigma_j$, because the leads may possess different magnetic axes.

Similarly the thermal energy carried by the electrons from reservoir $i$  per unit time is
\begin{equation}
I^{i}_{q}=\int(dE/h) \sum_{\sigma_i, j,\sigma_j}(E-\mu_i)P^{i,j}_{\sigma_i,\sigma_j} \bar f^j_{\beta_j}(E)  \label{xxxx2}
\end{equation}
For the special case of a normal  multi-terminal junction having collinear magnetic moments, $\alpha_i = \sigma$ for all $i$ and since there is no spin-flip scattering, $P^{i,j}_{\sigma,\sigma'} = P^{i,j}_{\sigma,\sigma}\delta_{\sigma,\sigma'}$. 
In this case, the total Hamiltonian of the whole system is block diagonal in spin indices and the scattering matrix can be obtained from separate calculations for each spin.
I assume that initially the junction is in thermodynamic equilibrium, so that all reservoirs possess the same chemical potential $\mu_0$. Subsequently, I apply to each reservoir $i$ a different voltage $V_i$, so that its chemical potential is $\mu_i=\mu_0-e\,V_i$. Then from equation (\ref{xxxx22}), the charge per unit time per spin entering the scatterer from each lead can be written as
\begin{equation}
I^{i}_e=-e\int(dE/h) \sum_{\sigma, j}P^{i,j}_{\sigma,\sigma} \bar f^j_{\sigma}(E)  \label{xx3}
\end{equation}
and the thermal energy per spin per unit time is
\begin{equation}
I^{i}_q=\int(dE/h) \sum_{\sigma, j}(E-\mu_i)P^{i,j}_{\sigma,\sigma} \bar f^j_{\sigma}(E)  \label{x4}
\end{equation}
where  $e=\vert e \vert$ and $\bar f^i_\sigma(E)=f(E-\mu_i)-f(E-\mu)$ is the deviation in Fermi distribution of lead $i$ from the reference distribution $f(E-\mu)$.

In the linear-response regime, the electric current $I$ and heat current $\dot{Q}$ passing through a device is related to the voltage difference $\Delta V$ and temperature difference $\Delta T$ by

\begin{equation}
 \begin{pmatrix}
\Delta V \\ \dot{Q} 
\end{pmatrix}
=
\begin{pmatrix}
G^{-1} & -S\\ \Pi & \kappa_{el}
\end{pmatrix}
\begin{pmatrix}
I \\ \Delta T
\end{pmatrix}  \label{x004}
\end{equation}
where electrical conductance $G$ (thermal conductance $\kappa_{el}$) is the ability of the device to conduct electricity (heat) and the thermopower $S^e$ (Peltier $\Pi$) is a measure of generated voltage (temperature) due to a temperature (voltage) differences between two sides of the device. In the limit of small potential differences or small differences in reservoir temperatures, the deviations in the distributions from the reference distribution $ \bar f^j_{\sigma}(E)$ can be approximated by differentials and therefore to evaluate currents, in the presence of collinear magnetism, the following spin-dependent integrals provided
\begin{equation}
 L^n_{ij,\sigma}(T,E_F)=\int_{-\infty}^{\infty}\,dE\,(E-E_F)^n\,T^{ij}_{\sigma,\sigma}(E,E_F)\,\left(-\frac{\partial f}{\partial E}\right)
\end{equation}
where $f(E,T)=(1+e^{(E-E_F)/k_BT})^{-1}$ is Fermi-Dirac distribution function and $k_B$ is Boltzmanns constant. In the presence of two leads labeled $i=1,2$, the spin-dependent low-voltage electrical conductance $G(T,E_F)$, the thermopower (Seebeck coefficient) $S(T,E_F)$, the Peltier coefficient $\Pi(T,E_F)$ and the thermal conductance due to the electrons $\kappa_{el}(T,E_F)$ as a function of Fermi energy $E_F$ and temperature $T$ can be obtained as
\begin{eqnarray}
 G(T,E_F)&=&\sum_{\sigma}\frac{e^2}{h}\,L_{12,\sigma}^0\nonumber\\
 S(T,E_F)&=&-\frac{1}{e\,T}\,\frac{\sum_{\sigma}L^1_{12,\sigma}}{\sum_{\sigma}L^0_{12,\sigma}}\nonumber\\
 \Pi(T,E_F)&=&T\,S(T,E_F)\nonumber\\
 \kappa_{el}(T,E_F)&=&\frac{1}{h\,T}\left(\sum_{\sigma}L_{12,\sigma}^2-\frac{(\sum_{\sigma}L_{12,\sigma}^1)^2}{\sum_{\sigma}L_{12,\sigma}^0}\right)
\end{eqnarray}
Note that the thermal conductance is guaranteed to be positive, because the expectation value of the square of a variable is greater than or equal to the square of the expectation value. 

Efficency of a thermoelectric matrial $\eta$ is defined as the ratio between the work done per unit time against the chemical potential difference (between two hot and cold reservior) and the heat extracted from the hot reservior per unit time. The maximum efficiency $\eta_{max}$ could be written as:
\begin{equation}
\eta_{max} = \frac{\Delta T}{T_h}\frac{\sqrt{Z.T_{avg}+1}-1}{\sqrt{Z.T_{avg}+1}+\frac{T_c}{T_h}}   \label{TE0}
\end{equation}
where $T_h$ and $T_c$ are the hot- and cold-side temperatures, respectively, $\Delta T=T_h-T_c$ and $T_{avg}=(T_h+T_c)/2$. The thermoelectric conversion efficiency (equation \ref{TE0}) is the product of the Carnot efficiency ($\frac{\Delta T}{T_h}$) and a reduction factor as a function of the material’s figure of merit $Z = S^2G\kappa^{-1}$, where $S$, $G$, and $\kappa = \kappa_{el}+\kappa_{ph}$ are the Seebeck coefficient, electrical conductance, and thermal conductance due to both electrons and phonons, respectively. More commonly a dimensionless figure of merit ($ZT = Z.T_{avg}$) is used to account for the efficency of the thermoelectric materials. The thermoelectric figure of merit could be written as
\begin{equation}
ZT = ZT_{el} \frac{\kappa_{el}}{\kappa_{el}+\kappa_{ph}}
\end{equation}
where the electronic thermoelectric figure of merit for a two-terminal system is
\begin{equation}
ZT_{el}=\frac{L^1_{12}}{L^0_{12}\,L^2_{12}-L^1_{12}}
\end{equation}
To calculate the total $ZT$, not only the thermal conductance due to the electrons are needed but also it is absolutely crucial to take the phonons contribution to the thermal conductance ($\kappa_{ph}$) into account as described in the next section.

\subsection{Phonon thermal conductance}
To calculate the heat flux through a molecular junction carried by the phonons, the equation \ref{xxxx22} could be used where the thermal conductance due to the phonons $\kappa_{ph}$ could be obtained \cite{sadeghi2015oligoyne} by calculating the phononic transmission $T_{ph}$ for different vibrational modes as

\begin{equation}
 \kappa_{ph}(T)=\frac{1}{2\pi}\int_0^\infty \hbar\omega T_{ph}(\omega)\frac{\partial f_{BE}(\omega,T)}{\partial T}d\omega
\end{equation}
where $f_{BE}(\omega,T)=(e^{\hbar \omega/k_BT}-1)^{-1}$ is Bose-Einstein distribution function and $\hbar$ is reduced Planck’s constant and $k_B$ is Boltzmann’s constant. To calculate the vibrational modes of a system, I use the harmonic approximation method to construct the dynamical matrix $D$. From the ground state relaxed $xyz$ coordinate of the system, each atom is displaced from its equilibrium position by $\delta q’$ and $–\delta q’$ in $x$, $y$ and $z$ directions and the forces $F^q_i=(F^x_i,F^y_i,F^z_i)$ in three directions $q_i=(x_i,y_i,z_i)$ on each atoms calculated. For $3n$ degrees of freedom ($n$ = number of atoms), the $3n\times 3n$ dynamical matrix $D$ is constructed

\begin{equation}
D_{ij} = \frac{K_{ij}^{qq'}}{M_{ij}}
\end{equation}
where $K_{ij}^{qq'}$ for $i \neq j$ are obtained from finite differences

\begin{equation}
K_{ij}^{qq'}=\frac{F_i^q(\delta q'_j)-F_i^q(\delta q'_j)}{2\delta q'_j}
\end{equation}
and the mass matrix $M=\sqrt{M_iM_j}$. To satisfy momentum conservation, the $Ks$ for $i=j$ (diagonal terms) are calculated from $k_{ii}=-\sum_{i \neq j} K_{ij}$. Once the dynamical matrix is constructed the Green's function method as described in \ref{Generalized model to calculate T(E)} could be used to calculate the phononic transmission coefficent $T_{ph}$.

\subsection{Spectral adjustment}
Although DFT is good at predicting the trends, it usually underestimates the position of the Fermi energy $E_F$, the exact energy levels (Kohn-Sham eigenvalues \cite{Seminario1995}) and therefore the position of the HOMO and LUMO and the energy gap. Therefore, to compare mean-field theory with experiment, some corrections are needed. One way is to use hybrid functionals e.g. B3LYP \cite{becke1993new} or many body calculations e.g. GW approximation \cite{PhysRev139A796}. These methods are either computationally very expensive (GW) where you cannot do calculation for a system with about 100 atoms in the best supercomputers today or they are fitted parameters to the experiment where their accuracy is not definite in new structures. For example, B3LYP combines the Hartree potential which usually overestimates the energy gap within the Kohn-Sham scheme which usually underestimate it to give more realistic gap. An alternative way is to correct the HOMO-LUMO gap using the values measured experimentally. A phenomenological scheme that improves the agreement between theoretical simulations and experiments in, for example, single-molecule electronics consists of shifting the occupied and unoccupied levels of the M (e.g. Molecule) region downwards and upwards respectively to increase the energy gap of the M region. 
The procedure is conveniently called spectral adjustment in nanoscale transport (SAINT) \cite{ferrer2014gollum}. The Hamiltonian $K=H-ES$ of a given M region could be modified as:

\begin{equation}
K_\mathrm{M}=K_\mathrm{M}^0+(\Delta_\mathrm{o}-\Delta_\mathrm{u})\,S_\mathrm{M}\,
\rho_\mathrm{M}\,S_\mathrm{M}+\Delta_\mathrm{u}\,S_\mathrm{M}
\end{equation}
where $\Delta_\mathrm{o,u}$ are energy shifts and ($no$, $nu$) denote the occupied and unoccupied states, respectively. $\rho_\mathrm{M} = \sum_{no}\,|\Psi_{no}\rangle\langle\Psi_{no}|\nonumber$ is the density matrix and $S_M$ is overlap matrix. If experimental HOMO and LUMO energies are available, $\Delta_\mathrm{o,u}$ can be chosen to correct HOMO and LUMO obtained from mean-field Hamiltonian. Alternatively, in the simplest case, the shifts $\Delta_\mathrm{o,u}$ are chosen to align the highest occupied and lowest unoccupied molecular orbitals (ie the HOMO and LUMO) with (minus) the ionization potential (IP) and 
electron affinity (EA)
of the isolated molecule
\begin{eqnarray}
 \Delta_\mathrm{o}^0&=&\epsilon_\mathrm{HOMO}+IP\nonumber\\
 \Delta_\mathrm{u}^0&=&-(\epsilon_\mathrm{LUMO}+EA)
\end{eqnarray}
However the Coulomb interactions in the isolated molecule are screened if the molecule is placed in close proximity to the metallic electrodes. This could be taken into account by using a simple image charge model, where the molecule is replaced by a point charge located at the middle point of the molecule and where the image planes are placed {1 \AA} above the electrodes' surfaces. Then the shifts are corrected by screening effects $\Delta_\mathrm{o,u}=\Delta_\mathrm{o,u}^0+{e^2 \ln 2}/({8\,\pi\epsilon_0 a})$ where $a$ is the distance between the image plane and the point image charge.

\subsection{Inclusion of a Gauge field}
For a scattering region of area $A$, if a magnetic field $B$ is applied the magnetic flux $\phi = B \times A$. To compute transport properties in the presence of a magnetic field, a Peierls substitution could be introduced by changing the phase factors of the coupling elements between atomic orbitals. For example in the case of a nearest-neighbor tight-binding Hamiltonian, the hoping matrix element $H_{ij}$ between site $i$ and site $j$ is replaced with the modified element,
\begin{eqnarray}
H_{ij}^{B}=H_{ij}e^{-i\phi},
\end{eqnarray}
where 

\begin{equation}
\phi = \frac{e}{\hbar}\int_{\mathbf{r}_{j}}^{\mathbf{r}_{i}}\mathbf{A}(\mathbf{r})d\mathbf{r}
\end{equation}
and $\mathbf{r}_{i}$ and $\mathbf{r}_{j}$ are the positions of site $i$ and $j$ and $\mathbf{A}$ is the vector potential. The gauge should be chosen such that the principal layers of the leads remain translationally invariant after the substitution.

\subsection{Superconducting systems}
\begin{figure}
\centerline{\includegraphics[trim=1cm 1cm 1cm 1cm, width=0.5\columnwidth]{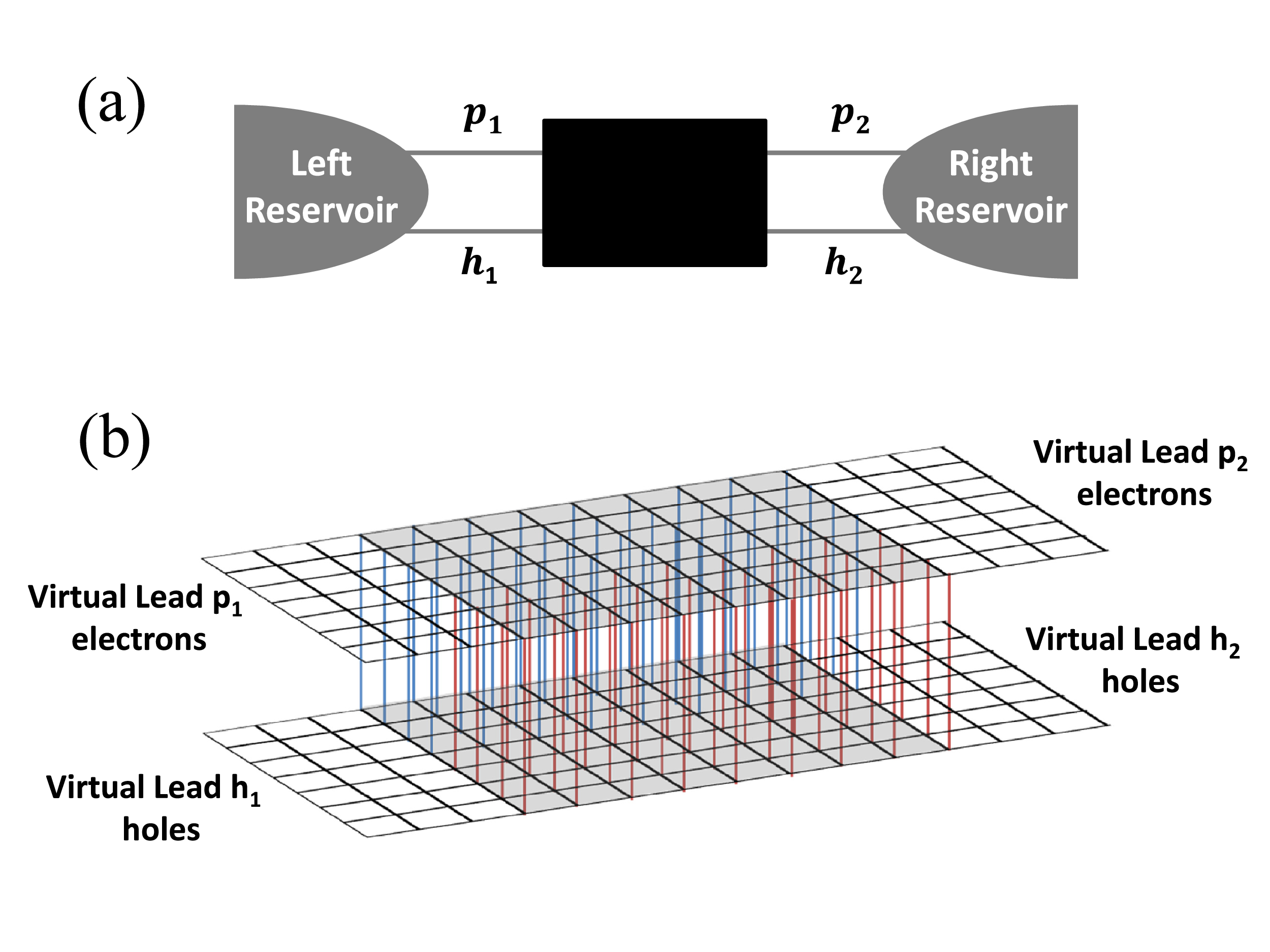}}
\caption{Two-probe device consist of reservoirs $\alpha$ and $\beta$ connected to a superconductor}
\label{fig:superconductor}
\end{figure}
Figure \ref{fig:superconductor}a shows a two-probe normal-superconductor-normal (N-S-N) device with left and right normal reservoirs connected to a scattering region containing one or more superconductors. If the complete Hamiltonian describing a normal system is  $H_N$, then in the presence of superconductivity within the extended scattering region, the new system is described by the Bogoliubov-de Gennes Hamiltonian
\begin{equation}
 H=\left(\begin{array}{cc}
          H_N&\Delta\\
          \Delta^*&-H_N^*\\
         \end{array}\right)\,\,\label{super}
\end{equation}
where the elements of the matrix $\Delta$ are non-zero only in the region occupied by a superconductor,  as indicated in figure 
\ref{fig:superconductor}b. Physically, $H_N$ describes particle degrees of freedom, $-H_N^*$ describes hole degrees of freedom and 
$\Delta$ is the superconducting order parameter.

The multi-channel scattering theory for such a normal-superconducting-normal (N-S-N) structure could be written as \cite{Lambert1998}:
\begin{equation}
 \left(\begin{array}{cc}
          I_{left}\\
          I_{right}\\
         \end{array}\right)
         =\frac{2\,e^2}{h}\,a\,\left(\begin{array}{cc}
          \frac{\mu_{left}-\mu}{e}\\
          \frac{\mu_{right}-\mu}{e}\\
         \end{array}\right)\,\,\label{x7}
\end{equation}
where $I_{left}$ ($I_{right}$) is the current from the left (right) reservoir, $\mu_{left}-\mu$ ($\mu_{right}-\mu$) is 
the difference between the chemical potential of the left (right) reservoir and the chemical potential $\mu$ of the 
superconducting condensate and the voltage difference between the left and right reservoirs is 
$(\mu_{left}-\mu_{right})/e$. In this equation,
\begin{equation}
 a=\left(\begin{array}{cc}
          N_{left}-R_o+R_a & -T_o'+T_a'\\
          -T_o+T_a& N_{right}-R_o'+R_a'\\
         \end{array}\right)\,\,
\end{equation}
where $N_{left}$ ($N_{right}$) is the number of open channels in the left (right) lead, $R_o, T_o$ ($R_a, T_a$) are normal (Andreev) reflection and transmission coefficients for quasi-particles emitted from the right lead, $R_o', T_o'$ ($R_a', T_a'$) are normal (Andreev) reflection and transmission coefficients from the left lead and all quantities are evaluated at the Fermi energy $E=\mu$. As a consequence of unitarity of the scattering matrix, these satisfy $R_o+ T_o+ R_a+ T_a= N_{left}$ and $R_o'+ T_o'+ R_a'+ T_a'= N_{right}$.

The current-voltage relation of Equ. (\ref{x7}) is fundamentally different from that encountered for normal systems, because 
unitarity of the s-matrix does not imply that the sum of each row or column of the matrix $a$ is zero. 
Consequently, the currents do not automatically depend solely of the applied voltage difference 
$(\mu_{left}-\mu_{right})/e$ (or more generally on the differences between incoming quasi-article 
distributions). In practice such a dependence arises only after the chemical potential of the 
superconductor adjusts itself self-consistently to ensure that the current from the left reservoir is equal to the 
current entering the right reservoir. Insisting that $I_{left}=-I_{right}=I$, the two-probe conductance $G=I/((\mu_{left}-\mu_{right})/e)$ takes the form of
\begin{equation}
G=\frac{2\,e^2}{h}\,\frac{a_{11}a_{22}-a_{12}a_{21}}{a_{11}+a_{22}+a_{12}+a_{21}}
\label{eq:Gsup}
\end{equation}
The above equation demonstrates why a superconductor possesses zero resistivity, because if the 
superconductor is disordered, then as the length $L$ of the superconductor increases, all transmission 
coefficients will vanish. In this limit, the above equation reduces to $(h/2e^2)G=2/R_a + 2/R_a'$. 
In contrast with a normal scatterer, this shows that in the presence of Andreev scattering, as $L$ 
tends to infinity, the resistance ( = 1/conductance) remains finite and therefore the resistivity (ie resistance per unit 
length) vanishes.

\subsection{Environmental effects}
To model environmental effects e.g. water, counter-ions, etc on the transport properties of a molecular junction, usually a statistical analysis needs to be carried out. Since a molecular junction in the presence of the surrounding molecules is a dynamic object at room temperature, a molecular dynamics simulation is usually needed first, to understand the range of possible configurations of the system. A few configuration then should be extracted and full DFT calculations carried out to obtain the mean field Hamiltonian of the system in the presence of the surrounding molecules. Another way to study the environmental effect is to create a series of configurations in the presence of the surrounding molecules in a more systematic but less physical way e.g. by moving the surroundings artificially in different directions. Then without geometry relaxation, one could find the binding energy of the surroundings to the backbone of the molecule for each configuration and only study those with higher binding energies. Both of these methods are widely used in the literature to model environmental effects. It is worth mentioning that since different effects such as physobrtion, charge transfer, etc could play important role in these simulations, SCF methods need to be used to calculated the transport from mean-field Hamiltonian.

\bibliographystyle{ieeetr}
\bibliography{newbiblo}
\end{document}